\documentclass[aps,prd,superscriptaddress,preprintnumbers,nofootinbib,10pt]{revtex4-2}
\usepackage{multirow}
\usepackage{amsmath}
\usepackage{amssymb}
\usepackage[dvipdf,dvips]{graphicx}
\usepackage{hyperref}
\usepackage{url}
\usepackage{slashed}
\usepackage{subfig}
\usepackage[usenames,dvipsnames]{xcolor}
\usepackage{amsmath}
\usepackage{amsfonts}
\usepackage{float}
\usepackage{amssymb}
\usepackage{epsfig}
\usepackage{graphics}
\usepackage{euscript}
\usepackage{slashed}
\usepackage{epstopdf}
\usepackage[utf8]{inputenc}
\allowdisplaybreaks
\usepackage[normalem]{ulem}
\usepackage{pifont}
\usepackage{dsfont}
\usepackage{graphicx}
\usepackage{latexsym,braket}
\usepackage{tikz-feynman}
\usepackage{tikz-cd}
\usepackage{easyReview}
\usepackage{cancel}
\usepackage[normalem]{ulem}
\usepackage{svg}
\usepackage{pslatex}
\usepackage{siunitx}
\usepackage{mathtools}
\mathtoolsset{showonlyrefs}		
\usepackage{makecell,tabularx}
\usepackage{ragged2e,nomencl}
\usepackage{slashed}
\usepackage{extpfeil}
\usepackage{etoolbox}
\usepackage{bm}
\usepackage{bbm}

\newcommand{\bvec}[1]{%
	\mathchoice
	{\,\reflectbox{$\displaystyle\vec{\reflectbox{\!$\displaystyle#1$}}$}}
	{\,\reflectbox{$\textstyle\vec{\reflectbox{\!$\textstyle#1$}}$}}
	{\,\reflectbox{$\scriptstyle\vec{\reflectbox{\!$\scriptstyle#1$}}$}}
	{\,\reflectbox{$\scriptscriptstyle\vec{\reflectbox{\!$\scriptscriptstyle#1$}}$}}
}

\newcommand{\DD}[1]{\!\mathcal{D}#1\,}

\makeatletter
\DeclareRobustCommand{\cev}[1]{%
	{\mathpalette\do@cev{#1}}%
}
\newcommand{\do@cev}[2]{%
	\vbox{\offinterlineskip
		\sbox\z@{$\m@th#1 x$}%
		\ialign{##\cr
			\hidewidth\reflectbox{$\m@th#1\vec{}\mkern4mu$}\hidewidth\cr
			\noalign{\kern-\ht\z@}
			$\m@th#1#2$\cr
		}%
	}%
}
\makeatother

\makeatletter
\newlength{\negph@wd}
\DeclareRobustCommand{\negphantom}[1]{%
  \ifmmode
    \mathpalette\negph@math{#1}%
  \else
    \negph@do{#1}%
  \fi
}
\newcommand{\negph@math}[2]{\negph@do{$\m@th#1#2$}}
\newcommand{\negph@do}[1]{%
  \settowidth{\negph@wd}{#1}%
  \hspace*{-\negph@wd}%
}
\makeatother

\begin{document}

\title{On the Casimir effect with mixed dynamical edge mode and\\perfect electromagnetic conducting boundary conditions}

\author{Jarne Devroe}
\email{jarne.m.devroe@gmail.com}
\affiliation{KU Leuven Campus Kortrijk -- Kulak, Department of Physics, Etienne Sabbelaan 53 bus 7657, 8500 Kortrijk, Belgium}

\author{David Dudal}
\email{david.dudal@kuleuven.be}
\affiliation{KU Leuven Campus Kortrijk -- Kulak, Department of Physics, Etienne Sabbelaan 53 bus 7657, 8500 Kortrijk, Belgium}

\author{Sebbe Stouten}
\email{sebbe.stouten@kuleuven.be}
\affiliation{KU Leuven Campus Kortrijk -- Kulak, Department of Physics, Etienne Sabbelaan 53 bus 7657, 8500 Kortrijk, Belgium}

\begin{abstract}
	We study the Casimir effect for a parallel plate setup with one plate with dynamical edge mode (DEM) boundary conditions, and one plate with perfect electromagnetic conductor (PEMC) boundary conditions.
	In order to restore BRST invariance, new edge fields are introduced on the DEM plate.
	We then lift the boundary conditions into the action using Lagrange multiplier fields, and integrate out the bulk fields to obtain a non-local effective boundary theory from which we compute the Casimir energy.
	The resulting Casimir force is identical to a PMC--PEMC setup, implying that, from the point of view of the Casimir effect, a DEM plate is equivalent to a PMC plate.
	We also include a detailed derivation of the general functional method used to compute the Casimir energy from the partition function.
\end{abstract}

\maketitle

\section{Introduction}

Dynamical edge mode boundary conditions have recently received a lot of attention in the literature \cite{ballDynamicalEdgeModes2024,BallpForms2025,BallYM2026}.
DEM conditions are a clean way to introduce edge modes into a system, which are crucial objects for understanding entanglement entropy \cite{Donnelly:2014fua,Donnelly:2015hxa, Donnelly:2016auv,Donnelly:2020teo}.
Indeed, edge modes offer a statistical interpretation for the contact term in the entanglement entropy \cite{KABAT1995281,FURSAEV1997697,Solodukhin}.
These DEM conditions (implicitly) break gauge invariance on the boundary, thereby elevating previously gauge degrees of freedom to physical `entanglement' degrees of freedom: the edge modes.

In the recent work \cite{canforaDynamicalEdgeModes2025}, the Casimir effect between two parallel plates with dynamical edge mode (DEM) boundary conditions was studied.
As suggested there, it would be interesting to also study the Casimir effect for a mixed setup of one DEM plate and one perfect electromagnetic conductor (PEMC) plate.
The interplay of the two types of boundary conditions could a priori have non-trivial consequences, and since this setup explicitly breaks reflection symmetry, a repulsive Casimir force is not ruled out \cite{kennethOppositesAttractTheorem2006a}.
In fact, we will indeed find a repulsive Casimir force for a certain region of parameter space.

In \cite{Rode:2017yqy,dudalCasimirEnergyPerfect2024}, the Casimir force in $(3+1)$ dimensions between two parallel PEMC plates with duality angles \(\theta^-\) and \(\theta^+\) was found to be
\begin{align}
	\mathcal{F}_{\text{Cas}}^\text{PEMC--PEMC}(L, \theta^+ - \theta^-) =-\frac{3}{8 \pi^2 L^4} \mathrm{Re} \, \mathrm{Li}_4(e^{2i(\theta^+-\theta^-)}),
	\label{eq: Casimir force PEMC PEMC}
\end{align}
which only depends on the duality invariant angle difference \(\theta^+-\theta^-\).
If one considers a reflection symmetric setup where both plates satisfy exactly the same {PEMC} boundary conditions, i.e.~with the same duality angles $\theta^- = \theta^+$, one retrieves the standard attractive Casimir force between two perfectly conducting plates
\begin{align}\label{eq:PEC-PEC}
	\mathcal{F}^\text{PEMC--PEMC}_{\text{Cas}}(L, 0) = -\frac{\pi^2}{240 L^4}.
\end{align}
In the recent study \cite{canforaDynamicalEdgeModes2025}, the Casimir force for two DEM plates was computed and found to be identical to the PEMC--PEMC case with \(\theta^+ = \theta^-\):
\begin{align}
	\mathcal{F}^{\text{DEM--DEM}}_{\text{Cas}}(L) = -\frac{\pi^2}{240 L^4}. \label{eq: Casimir force DEM DEM}
\end{align}
This result suggests that, in terms of the Casimir effect, there is a similarity between DEM plates and PEMC plates.
The goal of this paper is to better understand this relation, which we will do by studying a mixed DEM--PEMC setup.
We will show that, as far as the Casimir effect is concerned, a DEM plate behaves exactly like a PMC plate.

This paper is organized as follows.
We start by giving a detailed discussion of the broadly applicable functional method to compute the Casimir energy in Sec.~\ref{sec:functional-method}.
In Sec.~\ref{ssec: action + PEMC DEM} we define the specific boundary conditions, action and geometry we will consider.
Since the DEM boundary conditions locally break gauge invariance on the plate, we need to restore BRST symmetry after quantization.
We do so in Sec.~\ref{ssec: BRST invariance}.
Then we compute the Casimir energy in several steps.
In Sec.~\ref{sssec: bulk fields PEMC - DEM} we integrate out the bulk fields, leaving us with a non-local effective boundary theory.
At that point, we compute the Casimir energy from this effective boundary theory in Landau gauge in Sec.~\ref{sssec: edge fields PEMC - DEM Landau}, and in Coulomb gauge in Sec.~\ref{sssec: edge fields PEMC - DEM Coulomb}.
Of course, both computations yield identical results in the end, which we discuss in Sec.~\ref{sec:casimir}.
We summarize our findings in Sec.~\ref{sec:conclusion}

\section{Casimir energy from the functional integral}\label{sec:functional-method}

The vacuum energy and consequently the Casimir force can be retrieved from the functional integral. To show this, let us consider a general field theory with field $\varphi(x)$ and Minkowski Lagrangian density $\mathcal{L}_M[\varphi, \partial_{\mu}\varphi]$. In the Heisenberg picture, the time dependent operator $\hat{\varphi}_H(\tau, \vec{x})$ for this field is given by
\begin{align*}
	\hat{\varphi}_H(\tau, \vec{x}) = e^{iH\tau} \hat{\varphi}(0, \vec{x}) e^{-iH\tau},
\end{align*}
where $H$ is the full Hamiltonian of the theory. We will denote the eigenstates for this operator at time $\tau$ by $\left|\varphi_a, \tau\right\rangle$:
\begin{align*}
	\hat{\varphi}_H(\tau, \vec{x})\left|\varphi_a, \tau\right\rangle_H = \varphi_a(\vec{x})\left|\varphi_a, \tau\right\rangle_H.
\end{align*}
These are related to the eigenstates at time $\tau = 0$ by
\begin{align*}
	\left|\varphi_a, \tau\right\rangle_H = e^{iH\tau} \left|\varphi_a, 0\right\rangle_H.
\end{align*}
Here, $\tau$ serves as a label indicating that $\left|\varphi_a, \tau\right\rangle_H$ is an eigenstate of the Heisenberg picture operator $\hat{\varphi}_H(\tau, \vec{x})$ at time $\tau$. Note that $\left|\varphi_a, \tau\right\rangle_H$ is \emph{not} the result of time evolving the state $\left|\psi(0)\right\rangle_S = \left|\varphi_a,0\right\rangle_H$ from $\tau = 0$ to time $\tau$ in the Schrödinger picture. The time evolved Schrödinger picture state would have a different time dependence factor, namely $e^{-iH\tau}$:
\begin{align*}
	\left|\psi(\tau)\right\rangle_S = e^{-iH\tau} \left|\psi(0)\right\rangle_S = e^{-iH\tau}\left|\varphi_a,0\right\rangle_H.
\end{align*}

Using the functional integral formalism (e.g.~\cite{peskinIntroductionQuantumField1995,weinbergQuantumTheoryFields1995}), we can calculate the transition amplitude going from an initial Heisenberg state $\varphi_i(\vec{x})$ at time $\tau = \tau_i$ to a final state $\varphi_f(\vec{x})$ at time $\tau = \tau_f$ as
\begin{align}
	\begin{split}
		\langle \varphi_f, \tau_f | \varphi_i, \tau_i \rangle & = \langle\varphi_f|e^{-iH(\tau_f-\tau_i)}|\varphi_i \rangle\\
		& =	\int_{\varphi(\tau_i, \vec{x}) =\varphi_i(\vec{x})}^{\varphi(\tau_f, \vec{x})=\varphi_f(\vec{x})} \mathcal{D}\varphi \ \exp\left[i\int_{\tau_i}^{\tau_f} \mathrm{d}\tau \int \mathrm{d}^3 \vec{x} \, \mathcal{L}_M[\varphi, \partial_\mu \varphi]\right]. \label{eq: Minkowski functional integral}
	\end{split}
\end{align}
We used the short hand notation $\left|\varphi_a\right\rangle \coloneqq \left|\varphi_a,0\right\rangle_H$, and we have dropped the subscript $H$ since we will exclusively work in the Heisenberg picture from now on.

In the end we want to find the vacuum energy, so it will be useful to relate the functional integral \eqref{eq: Minkowski functional integral} to the vacuum transition amplitude, going from $\tau_i = -\infty$ to $\tau_f = +\infty$:
\begin{align*}
	\langle \Omega, \tau_f = +\infty|\Omega, \tau_i = -\infty\rangle = \lim_{\substack{\tau_i \to -\infty \\ \tau_f \to +\infty}}\langle \Omega|e^{-iH(\tau_f-\tau_i)} |\Omega \rangle.
\end{align*}
In this expression $|\Omega\rangle$ denotes the ground state of the full theory with energy $H|\Omega\rangle = E_{\Omega}|\Omega\rangle$, where the vacuum energy is denoted by $E_{\Omega}$. Writing the energy eigenstates as $H |n\rangle = E_n |n\rangle$, we can introduce the resolution of the identity $\mathbbm{1} = \sum_n |n\rangle \langle n|$ on the left hand side of equation \eqref{eq: Minkowski functional integral}
\begin{align*}
	\langle\varphi_f, \tau_f | \varphi_i, \tau_i \rangle  = \sum_n \langle\varphi_f|e^{-iH(\tau_f-\tau_i)}|n\rangle \langle n|\varphi_i \rangle = e^{-iE_{\Omega}(\tau_f-\tau_i)} \langle\varphi_f|\Omega\rangle \langle \Omega|\varphi_i \rangle + \sum_{n\neq\Omega} e^{-iE_n(\tau_f-\tau_i)} \langle\varphi_f|n\rangle \langle n|\varphi_i \rangle.
\end{align*}

Next, the time coordinate $\tau$ can be Wick rotated to Euclidean time $t = i\tau$. After the Wick rotation, the transition amplitude becomes
\begin{align*}
	\langle\varphi_f, t_f | \varphi_i, t_i \rangle & = e^{-E_{\Omega}(t_f-t_i)} \langle\varphi_f|\Omega\rangle \langle \Omega|\varphi_i \rangle + \sum_{n\neq\Omega} e^{-E_n(t_f-t_i)} \langle\varphi_i|n\rangle \langle n|\varphi_f \rangle.
\end{align*}
Since the vacuum energy is the lowest energy, the first term is the dominant term in the limit $t_f-t_i \rightarrow +\infty$, and we can drop the other terms. This gives
\begin{align*}
	\langle\varphi_f, t_f=+\infty | \varphi_i, t_i = -\infty \rangle & = \lim_{\substack{t_i \to -\infty \\ t_f \to +\infty}} e^{-E_{\Omega}(t_f-t_i)} \langle\varphi_i|\Omega\rangle \langle \Omega|\varphi_f \rangle.
\end{align*}

Combining this result with the Wick rotated version of the functional integral \eqref{eq: Minkowski functional integral} and using arbitrary initial and final states $\varphi_i$ and $\varphi_f$, we can write
\begin{align*}
	& \langle \Omega, t_f = +\infty|\Omega, t_i = -\infty\rangle = \lim_{\substack{t_i \to -\infty \\ t_f \to +\infty}} \langle \Omega|e^{-H(t_f-t_i)} |\Omega \rangle = \lim_{\substack{t_i \to -\infty \\ t_f \to +\infty}} e^{-E_{\Omega}(\tau_f-\tau_i)}\\
	& = \left(\langle\varphi_i|\Omega\rangle \langle \Omega|\varphi_f \rangle\right)^{-1}\langle\varphi_f, t_f=+\infty | \varphi_i, t_i = -\infty \rangle\\
	& = \left(\langle\varphi_i|\Omega\rangle \langle \Omega|\varphi_f \rangle\right)^{-1} \int_{\varphi(t_i, \vec{x})=\varphi_i(\vec{x})}^{\varphi(t_f, \vec{x})=\varphi_f(\vec{x})} \mathcal{D}\varphi \ \exp\left[-\int_{t_i =-\infty}^{t_f=+\infty} \mathrm{d}t \int \mathrm{d}^3 \vec{x} \, \mathcal{L}_E[\varphi, \partial_\mu \varphi]\right]\\
	& = \left(\langle\varphi_i|\Omega\rangle \langle \Omega|\varphi_f \rangle\right)^{-1} \int \mathcal{D}\varphi \ \exp\left[-\int \mathrm{d}t \int \mathrm{d}^3 \vec{x} \, \mathcal{L}_E[\varphi, \partial_\mu \varphi]\right]\\
	& = \left(\langle\varphi_i|\Omega\rangle \langle \Omega|\varphi_f \rangle\right)^{-1} Z.
\end{align*}
In the second to last step the boundaries for the functional integral were dropped because $\varphi_i$ and $\varphi_f$ are arbitrary, and we recognized the partition function \(Z\) without sources in the last step.
The partition function can therefore be related to the vacuum energy $E_\Omega$:
\begin{align}
	Z = \lim_{\substack{t_i \to -\infty \\ t_f \to +\infty}} \left(\langle\varphi_i|\Omega\rangle \langle \Omega|\varphi_f \rangle\right) e^{-E_{\Omega}(t_f-t_i)}. \label{eq: vacuum energy from functional integral}
\end{align}

In most calculations in quantum field theory, such as those involving correlators, contributions of the vacuum are often discarded as being physically unobservable. This is often done by setting the (infinite) vacuum energy to zero, for example through normal ordering of creation and annihilation operators in second quantization, or by normalizing the functional integral.
However, the introduction of two parallel plates makes the vacuum state dependent on the plate separation $L$.
Consequently, instead of having one vacuum state $|\Omega\rangle$, we have an infinite set of distinct vacuum states $\{|\Omega_L\rangle, L \in \mathbb{R}^+\}$ for different plate separations. Each vacuum state $|\Omega_L\rangle$ has its own vacuum energy $E_{\Omega}^{(L)}$, which in general can differ for different vacuum states. Therefore, setting the vacuum energy to zero not only neglects the vacuum energies themselves, but also the differences between them. What Casimir considered in his original work on the attractive force between two perfectly conducting plates \cite{casimirAttractionTwoPerfectly1948} was the difference between the vacuum energy $E_{\Omega}^{(L)}$ at plate separation $L$, and the vacuum energy with infinite plate separation ($L \rightarrow \infty$), corresponding to the vacuum energy of free space, $E_{\Omega}^{(\infty)}$. So in this spirit we define the Casimir energy as
\begin{align}
	E_{\text{Cas}}(L) = E_{\Omega}^{(L)} - E_{\Omega}^{(\infty)}, \label{eq: Casimir energy definition}
\end{align}
effectively setting the vacuum energy of free space to zero $E_{\Omega}^{(\infty)} = 0$. By first regulating the calculation of both $E_{\Omega}^{(L)}$ and $E_{\Omega}^{(\infty)}$, taking the difference of these vacuum energies and removing the regulator, Casimir was able to get a physically observable, finite value for $E_{\text{Cas}}(L)$.

Now, let $Z^{(L)}$ and $Z^{(\infty)}$ denote the partition function for the system of plates with finite plate separation $L$ and infinite plate separation respectively. Taking the ratio of these partition functions yields
\begin{align*}
	\frac{Z^{(L)}}{Z^{(\infty)}} & = \lim_{\substack{t_i \to -\infty \\ t_f \to +\infty}} \frac{\langle\varphi_i|\Omega_L\rangle \langle \Omega_L|\varphi_f \rangle}{\langle\varphi_i|\Omega_\infty\rangle \langle \Omega_\infty|\varphi_f \rangle} \cdot \frac{e^{-E_{\Omega}^{(L)}(t_f-t_i)}}{e^{-E_{\Omega}^{(\infty)}(t_f-t_i)}}\\
	& = \lim_{\substack{t_i \to -\infty \\ t_f \to +\infty}} \frac{\langle\varphi_i|\Omega_L\rangle \langle \Omega_L|\varphi_f \rangle}{\langle\varphi_i|\Omega_\infty\rangle \langle \Omega_\infty|\varphi_f \rangle} \exp\left[-E_{\text{Cas}}(L)(t_f-t_i)\right].
\end{align*}
Taking the logarithm and dividing by the infinite time extent $T \coloneqq t_f-t_i$ gives
\begin{align*}
	\lim_{T \rightarrow \infty} \frac{1}{T}\ln\left[\frac{Z^{(L)}}{Z^{(\infty)}}\right] = \lim_{T \rightarrow \infty} \left\{\frac{1}{T} \ln\left[\frac{\langle\varphi_i|\Omega_L\rangle \langle \Omega_L|\varphi_f \rangle}{\langle\varphi_i|\Omega_\infty\rangle \langle \Omega_\infty|\varphi_f \rangle}\right] - E_{\text{Cas}}(L)\right\}.
\end{align*}
Assuming that the overlaps between the arbitrary initial/final states and the vacuum states, $\langle\varphi_{i/f}|\Omega_{L/\infty}\rangle$ are finite and nonzero, the first term on the right hand side will vanish in the infinite time extent limit, $T \rightarrow \infty$. Since the Casimir energy is independent of $T$, we have arrived at the expression relating the Casimir energy to the ratio of partition functions
\begin{align}
	E_{\text{Cas}}(L) = -\lim_{T \rightarrow \infty} \frac{1}{T}\ln\left[\frac{Z^{(L)}}{Z^{(\infty)}}\right]. \label{eq: Casimir energy and part function}
\end{align}

\section{Setup}\label{sec:setup}

Let us start with some notation conventions.
We will use the 4-vector notation \(v=(v_t,v_x,v_y,v_z)\), 3-vectors \(\cev v = (v_t,v_x,v_y)\) and \(\vec v = (v_x,v_y,v_z)\), and 2-vector \(\bar v=(v_x,v_y)\).
The corresponding index notations are Greek indices \(\mu,\nu,\dots\in\{t,x,y,z\}\), Latin indices \(i,j,\dots\in\{t,x,y\}\), Latin indices \(m,n,\dots\in\{x,y,z\}\), and Latin indices \(a,b,\dots\in\{x,y\}\).

\subsection{Action and mixed PEMC--DEM boundary conditions}\label{ssec: action + PEMC DEM}
In this paper, we consider the gauge fixed Maxwell action (i.e.~QED without fermions) in 4D Euclidean space. Its action is given by
\begin{align*}
	S & = S_{\text{Maxwell}} + S_{\text{GF}} = \int \mathrm{d}^4x \left\{\frac{1}{4}F_{\mu\nu}F_{\mu\nu} + \left( h \mathcal{F}[A]+\frac{\xi}{2}h^2\right) \right\},
\end{align*}
where $h$ is the Nakanishi-Lautrup field, \(\xi\) is the gauge parameter, and we denote the general gauge fixing condition with $\mathcal{F}[A]$. Later on, we will specialize to Landau gauge $\mathcal{F}_{\mathtt{L}}[A] = \partial_\mu A_\mu$ or Coulomb gauge $~{\mathcal{F}_{\mathtt{C}}[A] = \partial_m A_m}$.

\begin{figure}
	\centering
	\includegraphics[width=0.5\textwidth]{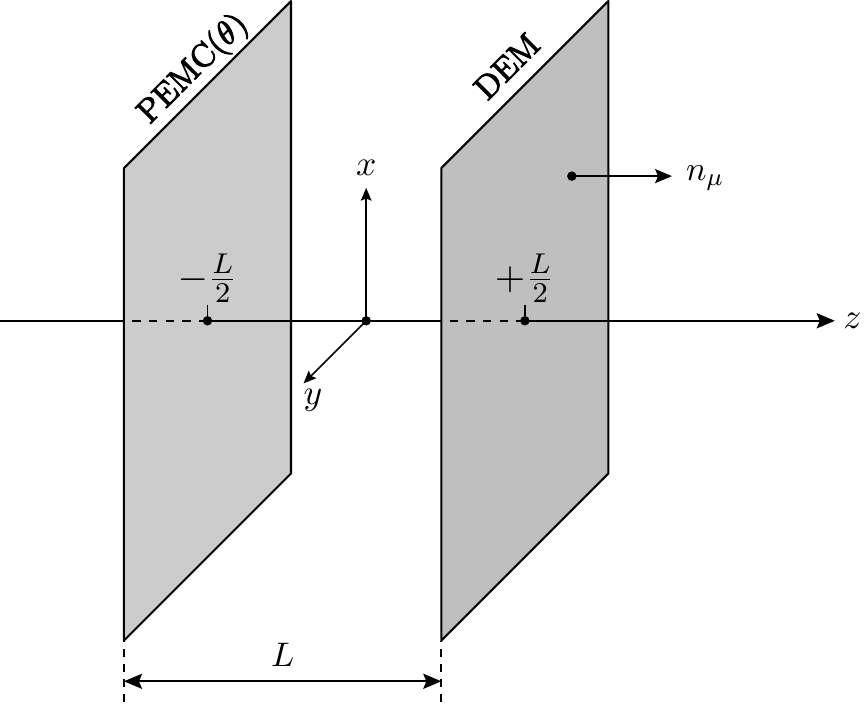}
	\caption{Representation of two infinitely large, infinitely thin parallel plates: the plate at $z=-\frac{L}{2}$ satisfies {PEMC} boundary conditions with duality angle $\theta$, while the plate at $z=+\frac{L}{2}$ satisfies {DEM} boundary conditions. The normal direction to the plates is given by the normal vector $n_\mu = \delta_{\mu z} = (0,0,0,1)$.}
	\label{fig: Casimir PEMC-DEM plate set up}
\end{figure}

The boundary configuration we will study consists of two infinitely large, infinitely thin, parallel plates. The plate located at \(z=z^-\coloneqq -L/2\) satisfies {PEMC} boundary conditions and the plate at \(z=z^+\coloneqq +L/2\) obeys {DEM} conditions, see Fig.~\ref{fig: Casimir PEMC-DEM plate set up}.
If we denote the normal vector on both plates by $n_\mu = (0,0,0,1)$, these boundary conditions are defined by
\begin{align}
	\text{PEMC:} \quad n_\mu G_{\mu\nu}(\theta) \coloneqq n_\mu \left(\cos(\theta) F_{\mu\nu} + \sin(\theta) \tilde{F}_{\mu\nu}\right) = 0\qquad\text{at }z=z^-,
	\label{eq: PEMC BC}
\end{align}
and
\begin{align}
	\text{{DEM}:} \quad
	\begin{cases}
		A_t = 0,\\
		n_\mu F_{\mu m} = 0\qquad\text{at }z=z^+.
	\end{cases}
	\label{eq: DEM conditions}
\end{align}
The {PEMC} conditions \cite{lindell2005perfect} are the most general linear gauge invariant boundary conditions, and contain the more common {PMC} conditions
$n_\mu F_{\mu\nu} = 0,$
and {PEC} conditions
$n_\mu \tilde F_{\mu\nu} = 0$
as special cases, resp.~by setting the duality angle \(\theta=0\) and \(\theta=\pi/2\).
Also note that {DEM} conditions are a combination of {PMC} and {PEC}.
Indeed: the $t$-direction of the {DEM} conditions implies {PEC} in that direction, whereas the $\{x,y,z\}$ directions are given by {PMC} conditions.

Clearly, the {PEMC} boundary conditions are gauge invariant, but the {DEM} boundary conditions are not: they break gauge invariance on the {DEM} plate located at $z = z^+$. The remaining gauge freedom of the theory are gauge transformations \mbox{$A_\mu \rightarrow A'_\mu = A_\mu + \partial_\mu \chi$} with vanishing support on \(z=z^+\), called ``small gauge transformations"; and gauge transformations $A_\mu \rightarrow A'_\mu = A_\mu + \partial_\mu \chi$ with non-vanishing support on \(z=z^+\), known as ``large gauge transformations", that satisfy
\begin{equation}\label{eq:large-dem-gauge-transformation}
	\partial_{t} \chi\big\vert_{z = z^+} = 0.
\end{equation}
Other large gauge transformations, with $\partial_{t} \chi \vert_{z = z^+} \neq 0$, will violate the {DEM} conditions. The fact that we lost certain large gauge symmetries will give rise to a new degree of freedom living on the plate. To understand this, let us for a moment consider working in the Coulomb gauge $\mathcal{F}_{\mathtt{C}}[A] = \partial_m A_m = 0$. In this gauge, the $t$-component of Maxwell's equation becomes,
\begin{align*}
	0 = \partial_\mu F_{\mu t} = \partial_m F_{m t} = {\vec{\partial}}^{\,2} A_t - \partial_t \partial_m A_m = {\vec{\partial}}^{\,2} A_t.
\end{align*}
Fourier transforming this equality yields, $\vec{k}^2 A_t = 0$ and as a result of this $A_t = 0$, such that we only need to consider $A_m$ in Coulomb gauge. As in \cite{ballDynamicalEdgeModes2024}, we can now decompose the photon field $A_m$ in Coulomb gauge as
\begin{align}
	A_m(t,\vec{x}) = \tilde{A}_m(t,\vec{x}) + \partial_m \alpha(t,\vec{x}) \label{eq: decomp photon field},
\end{align}
where $\tilde{A}$ satisfies $\partial_m \tilde{A}_m = 0$ and $n_m \tilde{A}_m|_{z=z^+} = 0$. Since we work in Coulomb gauge, this implies that the scalar field $\alpha(t,\vec{x})$ must fulfil ${\vec{\partial}}^{\,2} \alpha(t,\vec{x}) = 0$. Remark that under the Coulomb gauge, we actually still have some gauge redundancy left for functions $\chi$ solving ${\vec{\partial}}^{\,2} \chi = 0$ and $\partial_{t} \chi\big\vert_{z=z^+} = 0$.
As a consequence, as long as \(z\neq z^+\), we can gauge away the second part of decomposition \eqref{eq: decomp photon field}, with as result that
\begin{align*}
	A_m(t,\vec{x}) = \tilde{A}_m(t,\vec{x}) + \partial_m \alpha(t,\vec{x}) \xrightarrow{A_\mu \rightarrow A_\mu + \partial_\mu \chi} \tilde{A}_m(t,\vec{x}) + \partial_m \alpha(t,\vec{x}) + \partial_m \chi(t,\vec{x}) = \tilde{A}_m(t,\vec{x}),
\end{align*}
by choosing $\chi(t,\vec{x}) = -\alpha(t,\vec{x})$. On the {DEM} plate at \(z=z^+\), this is not possible, since there the gauge parameter $\chi$ cannot depend on time, cfr.~\eqref{eq:large-dem-gauge-transformation}.
Consequently, we can only shift away the second component in \eqref{eq: decomp photon field} at a fixed instant in time, but not for all of time. Hence the photon field in Coulomb gauge takes the form
\begin{align}
	A_m(t,\vec{x}) = \tilde{A}_m(t,\vec{x}) + \partial_m \alpha(\bvec{x}) \, \delta(z-z^+).
	\label{eq: DEM decomp}
\end{align}
Here the scalar field $\alpha(\bvec{x}) \coloneqq \alpha(x)\vert_{z = z^+}$ is a new field that solely lives on the {DEM} plate, and is thus referred to as an edge mode. In \cite{ballDynamicalEdgeModes2024,canforaDynamicalEdgeModes2025} it was shown that the conjugate field to this edge mode is given by the normal electrical field $E_\bot$ on the {DEM} plate. Together, $(E_\bot(\bvec{x}), \alpha(\bvec{x}))$ constitute the edge modes living on the {DEM} boundary. How these fields enter the action in the {PEMC}--{DEM} setup will become clear in the next section, when we will modify the action to ensure BRST invariance.

After this brief digression explaining how edge modes follow from {DEM} boundary conditions, let us now lift the boundary conditions into the action using Lagrange multiplier fields $b_\mu^\pm(\bvec{x})$:
\begin{align*}
	\text{{PEMC}:}  \quad \delta\!\left[n_\mu G_{\mu\nu}(\theta)\right]\!\Big|_{z= z^-} &= \int \DD{ b^-} \exp\left[-\!\int \mathrm{d}^4 x \,  n_\mu G_{\mu\nu}(\theta) b_\nu^-(\bvec{x}) \delta(z-z^-)\right]\\
	\text{{DEM}:}  \quad \delta\!\left[A_t\right]\delta\!\left[n_\mu F_{\mu m}\right]\!\Big|_{z= z^+} &= \int \DD{ b^+} \exp\left[-\!\int \mathrm{d}^4 x \left(A_t b_t^+(\bvec{x}) + n_\mu F_{\mu m} b_m^+(\bvec{x})\right)\delta(z-z^+)\,\right].
\end{align*}
Grouping both boundary conditions together, we get
\begin{align*}
	S_\text{BC} = \int \mathrm{d}^4x \left\{G_{zi}(\theta)b_i^-(\bvec{x}) \delta(z-z^-) + \left(A_t b_t^+(\bvec{x}) + F_{za} b_a^+(\bvec{x})\right)\delta(z-z^+)\right\}.
\end{align*}
After adding the boundary conditions for the two parallel plates to the action, the partition function thus reads
\begin{align*}
	Z = \int \DD{A} \DD{ b^+} \DD{b^-} e^{-\left( S_{\text{Maxwell}} + S_{\text{GF}} + S_{\text{BC}} \right)}.
\end{align*}

\subsection{Restoring BRST-invariance}\label{ssec: BRST invariance}
Since a computation of the Casimir energy has to yield a gauge invariant result, we first have to take into proper account the gauge invariance of the action on the {DEM} plate.
To address this issue, we need to take into account the ghost fields when quantizing our gauge theory in a spacetime with spatial boundaries \cite{mossBRSTInvariantBoundary1997,vassilevichFaddeevPopovTrickPresence1998,acharyyaBRSTSymmetryBoundary2016}.
The ghost term we need to add to the action is given by
\begin{align*}
	S_{\text{gh}} = \int \mathrm{d}^4x \, \bar{c}(x) \frac{\delta \mathcal{F}[A]}{\delta A_\mu} \partial_\mu c(x).
\end{align*}

After quantization, we should now require the quantum action to be BRST-invariant \cite{becchiAbelianHiggsKibbleModel1974,becchiRenormalizationAbelianHiggsKibble1975,becchiRenormalizationGaugeTheories1976,tyutinGaugeInvarianceField1975}.
Recall that the BRST transformation operator $s$ anticommutes with Grassmann valued fields such as $c$ and $\bar{c}$, and that it acts on the fields as
\begin{align*}
	sA_\mu &= -\partial_\mu c & s\bar{c} &= h & sb_i^\pm &= 0.\\*
	sc &= 0 & sh &= 0
\end{align*}
From the above transformation rules, it can immediately be observed that the BRST transformation operator is nilpotent, $s^2=0$.
Let us now inspect the BRST invariance of the action \(S_{\text{Maxwell}} + S_{\text{GF}} + S_{\text{BC}} + S_\text{gh}\).
Since $sF_{\mu\nu}=0$, we immediately have $s\left(\frac{1}{4}F_{\mu\nu}F_{\mu\nu}\right)=0$, $s\left(G_{zi}(\theta)b_i^-\right) = 0$, and $s\left(F_{za}b_a^+\right) = 0$. In addition, because of $s$ being nilpotent, we have that
\begin{align*}
	s\left(h\mathcal{F}[A] + \frac{\xi}{2}h^2 + \bar{c} \frac{\delta \mathcal{F}[A]}{\delta A_\mu} \partial_\mu c\right) = s^2\left(\bar{c} \frac{\delta \mathcal{F}[A]}{\delta A_\mu} A_\mu + \frac{\xi}{2}\bar{c}h\right) =0.
\end{align*}
(Here we assumed that we work in linear gauges for which $\frac{\delta \mathcal{F}[A]}{\delta A_\mu} A_\mu = \mathcal{F}[A]$.)
Therefore, the only action term breaking {BRST} invariance is the \(t\)-component of the {DEM} condition:
\begin{align*}
	\int \mathrm{d}^4x \ s(b_t^+ A_t)\delta(z-z^+) = -\int \mathrm{d}^4x \ (b_t^+ \partial_t c)\delta(z-z^+) \neq 0.
\end{align*}
To make the action fully BRST invariant, as in \cite{canforaDynamicalEdgeModes2025}, one has to introduce additional Grassmannian ghost fields $(\eta, \bar{\eta})$ and a scalar field $\gamma$, living on the {DEM} plate. These fields act as Lagrange multiplier fields to impose Dirichlet boundary conditions on the ghost fields $(c, \bar{c})$ and the Nakanishi-Lautrup field $h$ for the {DEM} plate, i.e., $c\vert_{z=z^{+}} = \bar{c}\vert_{z=z^{+}} = h\vert_{z=z^{+}} = 0$. Adding these Dirichlet boundary conditions as constraints to the action gives rise to the following new term
\begin{align*}
	S_{\text{ghBC}} = \int \mathrm{d}^4x \ \bigg\{\!\left(\bar{\eta}(\bvec{x})c + \eta(\bvec{x})\bar{c} + \gamma(\bvec{x})h\right) \delta(z-z^+)\bigg\}.
\end{align*}
It is worth noting that no additional boundary conditions for the ghost fields $c, \bar{c}$ and Nakanishi-Lautrup field $h$ are enforced on the $z = z^-$ plate with {PEMC} conditions.

With these new fields introduced, the extended set of BRST transformation rules is given by
\begin{align*}
	sA_\mu &= -\partial_\mu c & s\bar{c} &= h & s\gamma &= \eta & s\bar{\eta} &= -\partial_t b^+_t & sb_i^- &= 0\\
	sc &= 0 & sh &= 0 & s\eta &= 0 & sb_t^+ &= 0 & sb_a^+ &= 0.
\end{align*}
Again, by gathering BRST doublets in columns, it is clear that the extended BRST transformation is nilpotent as well, $s^2=0$. Notice that the last two terms in $S_{\text{ghBC}}$ can be written as a BRST exact term, namely $\eta \bar{c} + \gamma h = s(\gamma \bar{c})$, such that they are BRST invariant.
The BRST transformation of the extended action thus becomes
\begin{align*}
	sS & = s(S_{\text{Maxwell}} + S_{\text{GF}} + S_{\text{gh}} + S_{\text{BC}} + S_{\text{ghBC}})\\
	& = \int \mathrm{d}^4x \ s\left(b_t^+ A_t + \bar{\eta} c\right)\delta(z-z^+)
	 = -\int \mathrm{d}^4x \ \left(b_t^+ \partial_t c + \partial_t b_t^+ c\right)\delta(z-z^+)\\
	& = -\int \mathrm{d}^4x \ \partial_t(b_t^+ c) \delta(z-z^+) = 0,
\end{align*}\\
where we have used that the fields vanish at temporal infinity. We have therefore checked that the action \(S\) is BRST invariant. For completeness, let us gather the full action:
\begin{align}
	\begin{split}
		S & = S_{\text{Maxwell}} + S_{\text{GF}} + S_{\text{gh}} + S_{\text{BC}} + S_{\text{ghBC}}\\
		& = \int \mathrm{d}^4x \, \bigg\{ \frac{1}{4}F_{\mu\nu}F_{\mu\nu}
		+ h \mathcal{F}[A]+\frac{\xi}{2}h^2
		+ \bar{c}(x) \frac{\delta \mathcal{F}[A]}{\delta A_\mu} \partial_\mu c(x)\\
		& \qquad \qquad +  G_{zi}(\theta)b_i^-(\bvec{x}) \delta(z-z^-) + \left(A_t b_t^+(\bvec{x}) + F_{za} b_a^+(\bvec{x})\right)\delta(z-z^+) + \left(\bar{\eta}(\bvec{x})c + \eta(\bvec{x})\bar{c} + \gamma(\bvec{x})h\right) \delta(z-z^+)\bigg\}.
	\end{split}
	\label{eq: PEMC - DEM action after BRST}
\end{align}
We recall that in \cite{canforaDynamicalEdgeModes2025}, it was explicitly shown that the Lagrange multiplier field $b_t^+(\bvec{x})$ enforcing the $t$-component of the {DEM} condition, together with the Lagrange multiplier field $\gamma(\bvec{x})$ imposing Dirichlet boundary conditions for the Nakanishi-Lautrup field on the {DEM plate}, make up pairs of conjugate fields, corresponding to the original edge modes \((E_\bot(\bvec{x}),\alpha(\bvec{x}))\) introduced in \cite{ballDynamicalEdgeModes2024}, see Eq.~\eqref{eq: DEM decomp}.

\section{Casimir Energy PEMC--DEM}\label{sec: Casimir energy PEMC DEM}

Having constructed a fully BRST-invariant action which incorporates two infinite parallel plates where one satisfies {PEMC} and the other {DEM} boundary conditions, we are now ready to perform the calculation of the Casimir energy between these plates.
We will make use of the functional integral method described in Section \ref{sec:functional-method} to determine the Casimir energy.

Seeing that the Grassmann valued fields $(c, \bar{c}), (\eta,\bar{\eta})$ fully decouple from the other fields in action \eqref{eq: PEMC - DEM action after BRST}, we can factorize the partition function as
\begin{align*}
	Z^{(L)} & = \int \DD{A}\DD{b^+}\DD{b^-}\DD{c}\DD{\bar{c}}\DD{h}\DD{\eta}\DD{\bar{\eta}}\DD{\gamma} \ e^{-S[A, b^+, b^-, c, \bar{c}, h, \eta, \bar{\eta}, \gamma]}\\
	& = Z_{c,\bar{c},\eta, \bar{\eta}} \ Z^{(L)}_{A, b^+, b^-, h, \gamma}
\end{align*}
with
\begin{align*}
	&Z_{c,\bar{c},\eta, \bar{\eta}} = \int \DD{c}\DD{\bar{c}}\DD{\eta}\DD{\bar{\eta}} \exp\left[\int \mathrm{d}^4x \, \left\{\bar{c} \frac{\delta \mathcal{F}[A]}{\delta A_\mu} \partial_\mu c - \left(\bar{\eta}(\bvec{x})c + \eta(\bvec{x})\bar{c}\right) \delta(z-z^+)\right\}\right],\\
	&Z^{(L)}_{A, b^+, b^-, h, \gamma} = \int \DD{A}\DD{b^+}\DD{b^-}\DD{h}\DD{\gamma} \ \exp\bigg[-\int \mathrm{d}^4x \, \bigg\{\frac{1}{4}F_{\mu\nu}F_{\mu\nu} - \left(h \mathcal{F}[A]+\frac{\xi}{2}h^2\right)\\
	&\qquad \qquad \quad \quad + \, G_{zi}(\theta)b_i^-(\bvec{x}) \delta(z-z^-) + \left(A_t b_t^+(\bvec{x}) + F_{za} b_a^+(\bvec{x})\right)\delta(z-z^+) + \gamma(\bvec{x})h \delta(z-z^+)\bigg\}\bigg].
\end{align*}
Notice that \(Z_{c,\bar{c},\eta, \bar{\eta}}\) only depends on \(z^+\), not \(z^-\). This implies that it is \(L\)-independent and will not contribute to the Casimir energy.
Indeed, using the expression \eqref{eq: Casimir energy and part function} relating the partition function with the Casimir energy, one finds
\begin{align}
	\begin{split}
		E_{\text{Cas}}(L) & = -\lim_{T \rightarrow \infty} \frac{1}{T}\ln\left[\frac{Z^{(L)}}{Z^{(\infty)}}\right] = \underbrace{-\lim_{T \rightarrow \infty} \frac{1}{T}\ln\left[\frac{Z_{c,\bar{c},\eta,\bar{\eta}}}{Z_{c,\bar{c},\eta,\bar{\eta}}}\right]}_{=0} -\lim_{T \rightarrow \infty} \frac{1}{T}\ln\left[\frac{Z^{(L)}_{A, b^+, b^-, h, \gamma}}{Z^{(\infty)}_{A, b^+, b^-, h, \gamma}}\right] = E_{\text{non-ghost}}(L).
	\end{split}
	\label{eq: Casimir energy split}
\end{align}
Note that the vanishing of the ghost part of the Casimir energy did not happen in the parallel plate setup with both plates satisfying {DEM} boundary conditions \cite{canforaDynamicalEdgeModes2025}. So unlike in \cite{canforaDynamicalEdgeModes2025}, we expect \(E_{\text{non-ghost}}(L)\) to be independent of the chosen gauge.

We will continue by first performing the functional integral over the bulk fields $h$ and $A_\mu$, followed by the functional integral over the plate fields $b_i^\pm(\bvec{x})$ and $\gamma(\bvec{x})$. For these computations, it will be necessary to choose a specific gauge condition $\mathcal{F}[A]$. We will work out in detail both the case of Landau gauge $\mathcal{F}_{\mathtt{L}}[A] = \partial_\mu A_\mu$ and Coulomb gauge $\mathcal{F}_{\mathtt{C}}[A] = \partial_m A_m$.

\subsection{Integrating out bulk fields}\label{sssec: bulk fields PEMC - DEM}
Let us first focus on the parts of the partition function $Z^{(L)}_{A, b^+, b^-, h, \gamma}$ that contain the Nakanishi-Lautrup field $h$,
\begin{align*}
	\int \DD{h} \exp \left[\int \mathrm{d}^4x \, \left\{h \mathcal{F}[A] + \frac{\xi}{2}h^2 - \gamma(\bvec{x})h \delta(z-z^+)\right\}\right].
\end{align*}
The \(h\)-field can readily be integrated out, yielding
\begin{align*}
	\exp\left[-\int \mathrm{d}^4x \ \frac{1}{2\xi}\Big[\mathcal{F}[A] - \gamma(\bvec{x}) \delta(z-z^+)\Big]^2\right].
\end{align*}
It can thus be seen that this factor actually imposes the gauge fixing condition $\mathcal{F}[A] - \gamma(\bvec{x}) \delta(z-z^+) = 0$ instead of the original $\mathcal{F}[A] = 0$. It turns out that by making the action BRST invariant, we have slightly modified the gauge fixing such that on the {DEM}-plate $\mathcal{F}[A]\vert_{z = z^+} = \gamma(\bvec{x})$, while in the rest of spacetime $\mathcal{F}[A] = 0$ still. Expanding the square in the integrand above, the exponential becomes
\begin{align*}
	\exp\left[- \int \mathrm{d}^4x \ \left\{\frac{1}{2\xi}\left(\mathcal{F}[A]\right)^2 - \frac{1}{\xi}\mathcal{F}[A]\gamma(\bvec{x}) \delta(z-z^+) + \frac{1}{2\xi}\gamma(\bvec{x}) \delta(z-z^+) \gamma(\bvec{x}) \delta(z-z^+)\right\}\right].
\end{align*}
Focusing on the last term in the exponent, working out the $z$-integral gives
\begin{align*}
	\int \mathrm{d}^4x \ \frac{1}{2\xi}\gamma(\bvec{x}) \delta(z-z^+) \gamma(\bvec{x}) \delta(z-z^+) = \delta(z^+-z^+) \int \mathrm{d}^3 \bvec{x} \ \frac{1}{2\xi} \gamma(\bvec{x}) \gamma(\bvec{x}).
\end{align*}
If we employ the method of dimensional regularization to regulate divergent integrals, it follows that $\delta(z^+-z^+) = \delta(0) = 0$ (see e.g. \cite[eq. (4.2.6)]{collinsRenormalizationIntroductionRenormalization1984}, \cite[page 63]{anselmiRenormalization2019} and \cite[below eq. (10.9)]{zinn-justinQuantumFieldTheory2021}). Hence the term quadratic in $\gamma$ vanishes in the action and does not contribute to the partition function.

After having completed the functional integral over $h$, the remaining action for the fields $A_\mu, b_i^\pm(\bvec{x})$ and $\gamma(\bvec{x})$ reads
\begin{align*}
	S[A, b^+, b^-, \gamma] & = \int \mathrm{d}^4x \ \bigg\{\frac{1}{4}F_{\mu\nu}F_{\mu\nu} +  \frac{1}{2\xi}\left(\mathcal{F}[A]\right)^2 - \frac{1}{\xi}\mathcal{F}[A]\gamma(\bvec{x}) \delta(z-z^+)\\
	& \qquad \qquad \qquad + G_{zi}(\theta)b_i^-(\bvec{x}) \delta(z-z^-) + \left(A_t b_t^+(\bvec{x}) + F_{za} b_a^+(\bvec{x})\right)\delta(z-z^+)\bigg\}.
\end{align*}
In order to integrate out $A_\mu$, we will complete the square, which is most easily done in Fourier space. The quadratic term in \(A\) takes the form
\begin{align*}
	\int \mathrm{d}^4x \ \left\{\frac{1}{4}F_{\mu\nu}F_{\mu\nu} +  \frac{1}{2\xi}\left(\mathcal{F}[A]\right)^2\right\} = \int \frac{\mathrm{d}^4k}{(2\pi)^4} \, \left\{\frac{1}{2} A_{\mu}(k)\mathbb{K}_{\mu\nu}A_{\nu}(-k)\right\}
\end{align*}
with the quadratic operator in the Landau gauge $\mathcal{F}_{\mathtt{L}}[A] = \partial_\mu A_\mu$ given by
\begin{align*}
	\mathbb{K}^{\mathtt{L}}_{\mu\nu} = \delta_{\mu\nu}k^2 - \left(1-\frac{1}{\xi}\right)k_\mu k_\nu
\end{align*}
and in the Coulomb gauge $\mathcal{F}_{\mathtt{C}}[A] = \partial_m A_m$ by
\begin{align*}
	\mathbb{K}^{\mathtt{C}}_{\mu\nu} = \delta_{\mu\nu}k^2 - k_\mu k_\nu + \frac{1}{\xi}\delta_{\mu m} \delta_{\nu n} k_m k_n.
\end{align*}
The Fourier transform of the {PEMC} term in the action can be written as
\begin{align*}
	\int \mathrm{d}^4x \ G_{zi}(\theta) b^-_i(\bvec{x}) \delta(z-z^-) =  \int \! \frac{\mathrm{d}^4k}{(2\pi)^4} \, A_\mu(k) w_\mu(-k)
\end{align*}
where we defined the vector $w_\mu(-k)$ as
\begin{align}
	\left\{
	\begin{aligned}
		w_i(-k) & = \left[-i \cos(\theta) k_z b^-_i(-\bvec{k}) + \varepsilon_{3ijk} \sin(\theta) k_j b^-_k(-\bvec{k})\right]e^{-i k_z z^-}\\
		w_z(-k) & = i \cos(\theta) k_i b^-_i(-\bvec{k})e^{-i k_z z^-}.
	\end{aligned}
	\right.
	\label{eq: PEMC source}
\end{align}
Next, Fourier transforming the term in the action responsible for enforcing the {DEM} boundary condition, one finds
\begin{align*}
	& \int \mathrm{d}^4 x \, \left[A_t b_t^+(\bvec{x}) +F_{za} b_a^+(\bvec{x})\right]\delta(z-z^+) = \int \mathrm{d}^4 x \, \left[A_t b_t^+(\bvec{x}) + 2\partial_{[a}A_{z]}b_a^+(\bvec{x})\right]\delta(z-z^+)\\
	& = \int \! \frac{\mathrm{d}^4k}{(2\pi)^4} \left[A_t(k) b_t^+(-\bvec{k})- 2ik_{[a}A_{z]}(k)b_a^+(-\bvec{k})\right] e^{-i k_z z^+} \eqqcolon \int \! \frac{\mathrm{d}^4k}{(2\pi)^4} \, A_\mu(k) m_\mu(-k)
\end{align*}
in which for notational convenience we introduced the vector $m_\mu(-k)$, defined as
\begin{align}
	\left\{
	\begin{aligned}
		m_t(-k) & = b_t^+(-\bvec{k}) e^{-i k_z z^+}\\
		m_a(-k) & = i k_z b_a^+(-\bvec{k}) e^{-i k_z z^+}\\
		m_z(-k) & = -i k_a b_a^+(-\bvec{k}) e^{-i k_z z^+}.
	\end{aligned}
	\right.
	\label{eq: DEM source}
\end{align}\\
The last term in the action that needs to be brought to Fourier space is the cross term $\mathcal{F}[A]\gamma(\bvec{x})$. Filling in the explicit expression for the Landau gauge fixing,  $\mathcal{F}_{\mathtt{L}}[A] = \partial_\mu A_\mu$ we find that in Fourier space
\begin{align*}
	& -\frac{1}{\xi} \int \mathrm{d}^4 x \, \mathcal{F}_{\mathtt{L}}[A]\gamma(\bvec{x}) \delta(z-z^+) = \frac{1}{\xi} \int \! \frac{\mathrm{d}^4k}{(2\pi)^4} i k_\mu A_\mu(k) \gamma(-\bvec{k}) e^{-i k_z z^+} = \int \! \frac{\mathrm{d}^4k}{(2\pi)^4} \, A_\mu(k) u^{\mathtt{L}}_\mu(-k)
\end{align*}
identifying the field $u^{\mathtt{L}}_\mu(-k)$ in the Landau gauge as
\begin{align}
	u^{\mathtt{L}}_\mu(-k) = \frac{i}{\xi} k_\mu \gamma(-\bvec{k}) e^{-i k_z z^+}.
	\label{eq: gamma Landau source}
\end{align}
Restricting the summation over the full set of spacetime indices $\mu \in \{x,y,z,t\}$ to only the spatial indices $m \in \{x,y,z\}$ in the above derivation, we find that in the Coulomb gauge $\mathcal{F}_{\mathtt{C}}[A] = \partial_m A_m$,
\begin{align*}
	& -\frac{1}{\xi} \int \mathrm{d}^4 x \, \mathcal{F}_{\mathtt{C}}[A]\gamma(\bvec{x}) \delta(z-z^+) = \int \! \frac{\mathrm{d}^4k}{(2\pi)^4} \, A_\mu(k) u^{\mathtt{C}}_\mu(-k)
\end{align*}
with field $u^{\mathtt{C}}_\mu(-k)$ now given by
\begin{align}
	u^{\mathtt{C}}_\mu(-k) = \frac{i}{\xi} \delta_{\mu m} k_m \gamma(-\bvec{k}) e^{-i k_z z^+}.
	\label{eq: gamma coulomb source}
\end{align}
In the above, to distinguish between the Landau and Coulomb gauge we added a superscript $\mathtt{L}$ or $\mathtt{C}$ where necessary.

Combining the above results we can write the action $S[A, b^+, b^-, \gamma]$ as a Gaussian with source term
\begin{align*}
	S[A, b^+, b^-, \gamma] = \int \frac{\mathrm{d}^4k}{(2\pi)^4} \, \left\{\frac{1}{2} A_{\mu}(k)\mathbb{K}_{\mu\nu}A_{\nu}(-k) + A_\mu(k)v_\mu(-k)\right\},
\end{align*}
where the source term $v_\mu(-k)$ now combines the terms from the {PEMC} and {DEM} plates with the cross term $\mathcal{F}[A]\gamma(\bvec{k})$ as
\begin{align}
	v_\mu(-k) \coloneqq w_\mu(-k) + m_\mu(-k) + u_\mu(-k).
	\label{eq: vsource PEMC DEM}
\end{align}
Completing the square and performing the Gaussian functional integral over $A_\mu$ will only contribute a multiplicative factor independent of the plate separation $L$ to the partition function. So moving forward we will neglect this factor since it will not contribute to the Casimir energy.

In the following subsections we will perform the functional integral over the plate fields $b_i^\pm(\bvec{k}), \gamma(\bvec{k})$
\begin{align*}
	Z^{(L)}_{b^+, b^-, \gamma} = \int \DD{ b^+} \DD{b^-}\DD{\gamma} \ \exp\left[\int \! \frac{\mathrm{d}^4k}{(2\pi)^4} \, \frac{1}{2}\left\{ v_\mu(k)(\mathbb{K}^{-1})_{\mu\nu}v_\nu(-k)\right\}\right]
\end{align*}
\noeqref{eq: PEMC source,eq: DEM source,eq: gamma Landau source,eq: gamma coulomb source,eq: vsource PEMC DEM}
in which the dependence on the Lagrange multiplier fields $b_i^\pm(\bvec{k})$ and $\gamma(\bvec{k})$ is encoded in the vector $v_\mu(-k)$, see definitions (\ref{eq: PEMC source}, \ref{eq: DEM source}, \ref{eq: gamma Landau source}, \ref{eq: gamma coulomb source}, \ref{eq: vsource PEMC DEM}). Since both the quadratic operator $\mathbb{K}$ and the source field $u_\mu(-k)$ depend on the choice of gauge fixing condition $\mathcal{F}[A]$ for their explicit expressions, we will first restrict our attention to the Landau gauge $\mathcal{F}_{\mathtt{L}}[A] = \partial_\mu A_\mu$ in \ref{sssec: edge fields PEMC - DEM Landau}. Afterwards, we will tackle the case of Coulomb gauge, $\mathcal{F}_{\mathtt{C}}[A] = \partial_m A_m$, in \ref{sssec: edge fields PEMC - DEM Coulomb}, focusing on how the calculation differs from the Landau case.

\subsection{Integrating out edge fields in Landau gauge}\label{sssec: edge fields PEMC - DEM Landau}
In Landau gauge, $\mathcal{F}_{\mathtt{L}}[A] = \partial_\mu A_\mu$, and we need to work with the functional integral
\begin{align}
	Z^{(L)}_{b^+, b^-, \gamma} = \int \DD{b^+}\DD{b^-}\DD{\gamma} \exp\left[\frac{1}{2} \int \! \frac{\mathrm{d}^4k}{(2\pi)^4} \, \left\{ v^{\mathtt{L}}_\mu(k)(\mathbb{K}^{\mathtt{L}})^{-1}_{\mu\nu}v^{\mathtt{L}}_\nu(-k)\right\}\right]
	\label{eq: part function Landau PEMC-DEM}
\end{align}
where the inverse of the quadratic operator now is given by
\begin{align*}
	(\mathbb{K}^{\mathtt{L}})^{-1}_{\mu\nu} = \frac{1}{k^2} \left({T}_{\mu\nu}(k) + \xi \, {L}_{\mu\nu}(k)\right) = \frac{1}{k^2}\delta_{\mu\nu} + \frac{\xi-1}{k^4} k_\mu k_\nu
\end{align*}
with projection operators ${L}_{\mu\nu}(k)\coloneqq k_\mu k_\nu/k^2$ and ${T}_{\mu\nu}(k)\coloneqq \delta_{\mu\nu}-{L}_{\mu\nu}(k)$. Before expanding the full action and making the dependence on $b_i^\pm$ and $\gamma$ explicit, it will prove useful to first consider the contractions of both projectors with the source fields $w_\mu(-k)$ and $u^{\mathtt{L}}_\mu(-k)$. It can easily be checked that
\begin{align*}
	w_\mu(k) {L}_{\mu\nu}(k) = 0, && w_\mu(k) {T}_{\mu\nu}(k) = w_\nu(k),
\end{align*}
and given that the source field $u^{\mathtt{L}}_\nu(-k)$ is proportional to the wavevector $k_\mu$, the action of the projection operators is simply given by
\begin{align*}
	u^{\mathtt{L}}_\mu(k) {T}_{\mu\nu}(k) = 0, && u^{\mathtt{L}}_\mu(k) {L}_{\mu\nu}(k) = u^{\mathtt{L}}_\nu(k).
\end{align*}
These results imply that the integrand in the action will reduce to
\begin{align}
	& v^{\mathtt{L}}_\mu(k)(\mathbb{K}^{\mathtt{L}})^{-1}_{\mu\nu}v^{\mathtt{L}}_\nu(-k) = \frac{1}{k^2} \, \Big\{v^{\mathtt{L}}_\mu(k) {T}_{\mu\nu}(k) v^{\mathtt{L}}_\nu(-k) + \xi v^{\mathtt{L}}_\mu(k){L}_{\mu\nu}(k) v^{\mathtt{L}}_\nu(-k) \Big\} \nonumber\\
	& = \frac{1}{k^2} \, \Big\{[w_\mu(k)+m_\mu(k)] {T}_{\mu\nu}(k) [w_\nu(-k)+m_\nu(-k)] \nonumber\\
	& \qquad \qquad + \xi [u^{\mathtt{L}}_\mu(k)+m_\mu(k)] {L}_{\mu\nu}(k) [u^{\mathtt{L}}_\nu(-k)+m_\nu(-k)] \Big\} \nonumber\\
	& = \frac{1}{k^2} \, \Big\{ w_\mu(k) w_\mu(-k) + w_\mu(k) m_\mu(-k) + m_\mu(k) w_\mu(-k) + (\xi-1) m_\mu(k) {L}_{\mu\nu}(k) m_\nu(-k) \nonumber\\
	& \qquad \qquad + \, m_\mu(k) m_\mu(-k) + \xi u^{\mathtt{L}}_\mu(k) u^{\mathtt{L}}_\mu(-k) + \xi u^{\mathtt{L}}_\mu(k) m_\mu(-k) + \xi m_\mu(k) u^{\mathtt{L}}_\mu(-k) \Big\}. \! \label{eq: integrand full PEMC-DEM}
\end{align}
Let us first focus on the terms quadratic in $b$. After some algebra, one finds for the terms with $w$ and $m$ that
\begin{align*}
	w_\mu(k) w_\mu(-k) & = b_i^-(-\bvec{k}) \left\{\cos^2(\theta)  \left[k_z^2 \delta_{ij} + k_i k_j\right] + \sin^2(\theta) \left[-\bvec{k}^2 \delta_{ij} + k_i k_j\right]\right\} b_j^-(\bvec{k})\\*
	m_\mu(k) m_\mu(-k) & = b_i^+(-\bvec{k}) \left\{\delta_{it}\delta_{tj} + \delta_{ia}\delta_{aj} k_z^2 +  \delta_{ia} k_a k_b \delta_{bj}\right\} b_j^+(\bvec{k})\\*
	m_\mu(k) L_{\mu\nu} m_\nu(-k) & = b_t^+(-\bvec{k}) \frac{k_t^2}{k^2} b_t^+(\bvec{k})\\
	w_\mu(k) m_\mu(-k) & = b_i^+(-\bvec{k}) \Big\{i \cos(\theta) \left[k_z \delta_{it} \delta_{tj} + ik_z^2 \delta_{ia} \delta_{aj} + i\delta_{ia} k_a k_j\right]\\*
	& \qquad \qquad \quad -\sin(\theta) \left[\varepsilon_{tzab} \delta_{it} k_a \delta_{bj} + i \varepsilon_{zakj}  \delta_{ia} k_k k_z\right]\Big\} e^{-i k_z L} b_j^-(\bvec{k})\\*
	m_\mu(k) w_\mu(-k) & = b_i^-(-\bvec{k}) \Big\{i \cos(\theta) \left[-k_z \delta_{it} \delta_{tj} + ik_z^2 \delta_{ia} \delta_{aj} + i k_i k_a \delta_{aj}\right]\\*
	& \qquad \qquad \quad -\sin(\theta) \left[-\varepsilon_{tzab} \delta_{ib} k_a \delta_{tj} + i \varepsilon_{zaki}  \delta_{aj} k_k k_z\right]\Big\} e^{i k_z L} b_j^+(\bvec{k}).
\end{align*}
Next we can find the term quadratic in $\gamma$ and the mixing terms $b \gamma$ from $u^{\mathtt{L}}$ and $m$ as
\begin{align*}
	u^{\mathtt{L}}_\mu(k) u^{\mathtt{L}}_\mu(-k) & = \frac{1}{\xi^2} k^2 \gamma(\bvec{k}) \gamma(-\bvec{k})\\
	m_\mu(k) u^{\mathtt{L}}_\mu(-k) & = \frac{i}{\xi} b_t^+(\bvec{k}) k_t \gamma(-\bvec{k})\\
	u^{\mathtt{L}}_\mu(k) m_\mu(-k) & = -\frac{i}{\xi} b_t^+(-\bvec{k}) k_t \gamma(\bvec{k}).
\end{align*}

As the subsequent step, these expressions are to be substituted into the integrand \eqref{eq: integrand full PEMC-DEM} and the $k_z$-integral in \eqref{eq: part function Landau PEMC-DEM} can be performed using the integral formulas in Appendix \ref{sec: useful fourier integrals}. The term quadratic in $\gamma$ does not contain any \(k_z\)-dependence, and is thus proportional to a Dirac delta:
\begin{align}
	\begin{split}
		S_{\gamma}^{\text{quad}} & = \frac{1}{2} \int \! \frac{\mathrm{d}^3\bvec{k}}{(2\pi)^3} \frac{1}{\xi} \gamma(\bvec{k}) \gamma(-\bvec{k}) \left(\int \! \frac{\mathrm{d}k_z}{2\pi}\right) = \frac{1}{2} \int \! \frac{\mathrm{d}^3\bvec{k}}{(2\pi)^3} \frac{1}{\xi} \gamma(\bvec{k}) \gamma(-\bvec{k}) \delta(0) = 0,
	\end{split}
	\label{eq: Sgammaquad}
\end{align}
which is zero under dimensional regularization, by the same argument as in the beginning of \ref{sssec: bulk fields PEMC - DEM}. Hence the partition function takes the Gaussian form
\begin{align}
	\!\!\! Z^{(L)}_{b^+, b^-, \gamma} \! = \! \int \! \DD{b^+}\DD{b^-}\DD{\gamma} \exp \! \left[-\frac{1}{2} \int \! \frac{\mathrm{d}^3\bvec{k}}{(2\pi)^3} \left\{b_i^\rho(-\bvec{k}) \mathbb{M}^{\rho\sigma}_{ij}(\bvec{k}) b_j^\sigma(\bvec{k}) + b_t^+(-\bvec{k}) s_{\mathtt{L}}(\bvec{k})\right\}\right]\!, \!\! \label{eq: part function gaussian with source Landau}
\end{align}
where upstairs indices \(\rho,\sigma,\dots\in \{-,+\}\) indicate the plate, and repeated plate indices are summed over.
The quadratic operator $\mathbb{M}^{\rho\sigma}_{ij}(\bvec{k})$ can be represented as a $6 \times 6$ matrix with $3 \times 3$ blocks given by
\begin{align*}
	\mathbb{M}^{++}_{ij}(\bvec{k}) & = -\frac{1}{4|\bvec{k}|^3}\left[(\xi-1)k_t^2 + 2 \bvec{k}^2\right] \delta_{it}\delta_{tj} + \frac{|\bvec{k}|}{2}\delta_{ia} \mathsf{T}_{\!ab}(\bvec{k}) \delta_{bj}\\*
	\mathbb{M}^{--}_{ij}(\bvec{k}) & = \frac{|\bvec{k}|}{2}\mathsf{T}_{\!ij}(\bvec{k}) - \lambda \mathsf{L}_{ij}(\bvec{k})\\
	\mathbb{M}^{+-}_{ij}(\bvec{k}) & = -\frac{|\bvec{k}|}{2} e^{-|\bvec{k}|L}\left\{\delta_{ia}\left[\mathsf{P}^1_{\!aj}(\bvec{k}) e^{-i\theta} + \mathsf{P}^2_{\!aj}(\bvec{k}) e^{i\theta}\right]\right. \\*
	& \left. \qquad \qquad \qquad \qquad + \frac{\delta_{it}}{|\bvec{k}|} \left[\mathsf{P}^1_{\!tj}(\bvec{k}) e^{-i\theta} + \mathsf{P}^2_{\!tj}(\bvec{k}) e^{i\theta} + \cos(\theta) \mathsf{L}_{tj}(\bvec{k})\right]\right\}\\*
	\mathbb{M}^{-+}_{ij}(\bvec{k}) & = -\frac{|\bvec{k}|}{2} e^{-|\bvec{k}|L}\left\{\delta_{aj}\left[\mathsf{P}^1_{\!ia}(\bvec{k}) e^{-i\theta} + \mathsf{P}^2_{\!ia}(\bvec{k}) e^{i\theta}\right]\right. \\*
	& \left. \qquad \qquad \qquad \qquad + \frac{\delta_{tj}}{|\bvec{k}|} \left[\mathsf{P}^1_{\!it}(\bvec{k}) e^{-i\theta} + \mathsf{P}^2_{\!it}(\bvec{k}) e^{i\theta} + \cos(\theta) \mathsf{L}_{it}(\bvec{k})\right]\right\}.
\end{align*}
The longitudinal and transverse projectors here are the 3-dimensional version:
\begin{align}
	\mathsf{L}_{ij}(\bvec{k}) \coloneqq \frac{k_i k_j}{\bvec{k}^2} && \text{ and } && \mathsf{T}_{\!ij}(\bvec{k}) \coloneqq \delta_{ij} - \frac{k_i k_j}{\bvec{k}^2} = \delta_{ij} - \mathsf{L}_{ij}(\bvec{k}) \label{eq: xyt projectors},
\intertext{and we have introduced two rank-1 projectors}
	\mathsf{P}^1_{\!ij}(\bvec{k}) \coloneqq \frac{1}{2}\left[\mathsf{T}_{\!ij}(\bvec{k}) + i \varepsilon_{zijk} \frac{k_k}{|\bvec{k}|}\right] && \text{ and } && \mathsf{P}^2_{\!ij}(\bvec{k}) \coloneqq \frac{1}{2}\left[\mathsf{T}_{\!ij}(\bvec{k}) - i \varepsilon_{zijk} \frac{k_k}{|\bvec{k}|}\right]. \label{eq: polarization projectors}
\end{align}
In the $\mathbb{M}^{--}_{ij}(\bvec{k})$ block matrix we already fixed the residual gauge symmetry $b_i^-(\cev{k})\rightarrow b_i^-(\cev{k}) + k_i \chi(\cev{k})$ by adding a gauge fixing term $-\lambda b_i^- \mathsf{L}_{ij}(\bvec{k}) b_j^-$.
From the above expressions for the quadratic operator $\mathbb{M}_{ij}^{\rho\sigma}(\bvec{k})$ it can be observed that $\mathbb{M}_{ij}^{\rho\sigma}(\bvec{k}) = \mathbb{M}_{ji}^{\sigma\rho}(-\bvec{k})$. The source term with field $s_{\mathtt{L}}(\bvec{k})$ in \eqref{eq: part function gaussian with source Landau} comes from the mixing term between $b$ and $\gamma$ and is given by
\begin{align}
	s_{\mathtt{L}}(\bvec{k}) = i \frac{k_t}{|\bvec{k}|} \gamma(\bvec{k}). \label{eq: source gamma Landau}
\end{align}

Rather than performing the functional integral with the matrix operator $\mathbb{M}(\bvec{k})$ given in the standard coordinate basis $(t,x,y)$, to simplify calculations, we choose to work in a real basis $(\textbf{e}_{\mathbf{1}}, \textbf{e}_{\mathbf{2}}, \textbf{e}_{\mathbf{t}})$ defined by
\begin{align*}
	\textbf{e}_{\mathbf{1}}(\bar{k}) & = \frac{1}{|\bar{k}|}(0,k_x, k_y) & \textbf{e}_{\mathbf{2}}(\bar{k}) & = \frac{1}{|\bar{k}|}(0,k_y, -k_x) & \textbf{e}_{\mathbf{t}} & = (1, 0, 0),
\end{align*}
where $\bar k = (k_x,k_y)$ and $|\bar{k}| = \sqrt{k_x^2 + k_y^2}$. Denoting the boundary fields $b^\rho(\bvec{k})$ in this new basis as $b_{\bm{\alpha}}^\rho(\bvec{k}) = (b_{\mathbf{1}}^-, b_{\mathbf{1}}^+, b_{\mathbf{2}}^-, b_{\mathbf{2}}^+, b_{\mathbf{t}}^-, b_{\mathbf{t}}^+)$ with indices $\alpha, \mathbf{\beta}$ running over $\{\mathbf{1}, \mathbf{2}, \mathbf{t}\}$, partition function \eqref{eq: part function gaussian with source Landau} becomes
\begin{align}
	\!\!\!\! Z^{(L)}_{b^+, b^-, \gamma} & = \! \int \DD{b^+}\DD{b^-}\DD{\gamma} \exp \left[-\frac{1}{2} \int \! \frac{\mathrm{d}^3\bvec{k}}{(2\pi)^3} \left\{b_{\bm{\alpha}}^\rho(-\bvec{k}) \mathbb{M}^{\rho\sigma}_{\bm{\alpha\beta}}(\bvec{k}) b_\mathbf{\bm{\beta}}^\sigma(\bvec{k}) + b_{\bm{\alpha}}^\rho(-\bvec{k}) j_{\bm{\alpha}}^\rho(\bvec{k})\right\}\right]\!. \!\! \label{eq: part function gaussian with source Landau basis2}
\end{align}
In the $(\textbf{e}_{\mathbf{1}}, \textbf{e}_{\mathbf{2}}, \textbf{e}_{\mathbf{t}})$ basis with $b_{\bm{\alpha}}^\rho(\bvec{k}) = (b_{\mathbf{1}}^-, b_{\mathbf{1}}^+, b_{\mathbf{2}}^-, b_{\mathbf{2}}^+, b_{\mathbf{t}}^-, b_{\mathbf{t}}^+)$ the quadratic operator is explicitly given by
\begin{align*}
	\!\!\!\!\!\!\!\!\!\!\!\!\!\!\!\!\!\!\!\!\!\!\!\!\!\!\!\!
	\mathbb{M}_{\bm{\alpha\beta}}^{\rho\sigma}(\bvec{k}) = \frac{1}{2\bvec{k}^2} \left(
	\begin{array}{cccccc}
		2 \bar{k}^2 \lambda - k_{t}^{2} |\bvec{k}| & k_{t}^{2} |\bvec{k}| e^{- L |\bvec{k}|} \cos{\left(\theta \right)} & 0 & - k_{t} \bvec{k}^2 e^{- L |\bvec{k}|} \sin{\left(\theta \right)} & |\bar{k}| k_{t} \left(2 \lambda + |\bvec{k}|\right) & 0\\
		k_{t}^{2} |\bvec{k}| e^{- L |\bvec{k}|} \cos{\left(\theta \right)} & - k_{t}^{2} |\bvec{k}| & - k_{t} \bvec{k}^2 e^{- L |\bvec{k}|} \sin{\left(\theta \right)} & 0 & - |\bar{k}| k_{t} |\bvec{k}| e^{- L |\bvec{k}|} \cos{\left(\theta \right)} & 0\\
		0 & k_{t} \bvec{k}^2 e^{- L |\bvec{k}|} \sin{\left(\theta \right)} & - |\bvec{k}|^{3} & |\bvec{k}|^{3} e^{- L |\bvec{k}|} \cos{\left(\theta \right)} & 0 & - |\bar{k}| |\bvec{k}| e^{- L |\bvec{k}|} \sin{\left(\theta \right)}\\
		k_{t} \bvec{k}^2 e^{- L |\bvec{k}|} \sin{\left(\theta \right)} & 0 & |\bvec{k}|^{3} e^{- L |\bvec{k}|} \cos{\left(\theta \right)} & - |\bvec{k}|^{3} & - |\bar{k}| \bvec{k}^2 e^{- L |\bvec{k}|} \sin{\left(\theta \right)} & 0\\
		- |\bar{k}| k_{t} \left(2 \lambda + |\bvec{k}|\right) & |\bar{k}| k_{t} |\bvec{k}| e^{- L |\bvec{k}|} \cos{\left(\theta \right)} & 0 & - |\bar{k}| \bvec{k}^2 e^{- L |\bvec{k}|} \sin{\left(\theta \right)} & \bar{k}^2 |\bvec{k}| - 2 \lambda k_{t}^{2} & - \bvec{k}^2 e^{- L |\bvec{k}|} \cos{\left(\theta \right)}\\
		0 & 0 & - |\bar{k}| |\bvec{k}| e^{- L |\bvec{k}|} \sin{\left(\theta \right)} & 0 & - \bvec{k}^2 e^{- L |\bvec{k}|} \cos{\left(\theta \right)} & \frac{- k_{t}^{2} \left(\xi - 1\right) - 2 \bvec{k}^2}{2 |\bvec{k}|}
	\end{array}
	\right),
\end{align*}
and for later use we identify the block matrices $\mathbf{A}, \mathbf{B}, \mathbf{C}, \mathbf{D}$ as
\begin{align}
	\mathbb{M}_{\bm{\alpha\beta}}^{\rho\sigma}(\bvec{k}) = \left(
	\begin{array}{cccccc}
		\mathbb{M}_{\mathbf{11}}^{--} & \mathbb{M}_{\mathbf{11}}^{-+} & \mathbb{M}_{\mathbf{12}}^{--} & \mathbb{M}_{\mathbf{12}}^{-+} & \mathbb{M}_{\mathbf{1t}}^{--} & \mathbb{M}_{\mathbf{1t}}^{-+}\\
		\mathbb{M}_{\mathbf{11}}^{+-} & \mathbb{M}_{\mathbf{11}}^{++} & \mathbb{M}_{\mathbf{12}}^{+-} & \mathbb{M}_{\mathbf{12}}^{++} & \mathbb{M}_{\mathbf{1t}}^{+-} & \mathbb{M}_{\mathbf{1t}}^{++}\\
		\mathbb{M}_{\mathbf{21}}^{--} & \mathbb{M}_{\mathbf{21}}^{-+} & \mathbb{M}_{\mathbf{22}}^{--} & \mathbb{M}_{\mathbf{22}}^{-+} & \mathbb{M}_{\mathbf{2t}}^{--} & \mathbb{M}_{\mathbf{2t}}^{-+}\\
		\mathbb{M}_{\mathbf{21}}^{+-} & \mathbb{M}_{\mathbf{21}}^{++} & \mathbb{M}_{\mathbf{22}}^{+-} & \mathbb{M}_{\mathbf{22}}^{++} & \mathbb{M}_{\mathbf{2t}}^{+-} & \mathbb{M}_{\mathbf{2t}}^{++}\\
		\mathbb{M}_{\mathbf{t1}}^{--} & \mathbb{M}_{\mathbf{t1}}^{-+} & \mathbb{M}_{\mathbf{t2}}^{--} & \mathbb{M}_{\mathbf{t2}}^{-+} & \mathbb{M}_{\mathbf{tt}}^{--} & \mathbb{M}_{\mathbf{tt}}^{-+}\\
		\mathbb{M}_{\mathbf{t1}}^{+-} & \mathbb{M}_{\mathbf{t1}}^{++} & \mathbb{M}_{\mathbf{t2}}^{+-} & \mathbb{M}_{\mathbf{t2}}^{++} & \mathbb{M}_{\mathbf{tt}}^{+-} & \mathbb{M}_{\mathbf{tt}}^{++}
	\end{array}
	\right)
	\eqqcolon \left(
	\begin{array}{cc}
		\mathbf{A}_{5\times5} & \mathbf{B}_{5\times1}\\
		\mathbf{C}_{1\times5} & \mathbf{D}_{1\times1}
	\end{array}
	\right).
	\label{eq: quadratic operator block matrix structure Landau}
\end{align}
In partition function \eqref{eq: part function gaussian with source Landau basis2} we also defined the source $j_{\bm{\alpha}}^\rho(\bvec{k})$ in terms of $s_{\mathtt{L}}(\bvec{k})$ as
\begin{align}
	j_{\bm{\alpha}}^\rho(\bvec{k}) = \delta_{\bm{\alpha}\mathbf{t}} \delta^{\rho+} s_{\mathtt{L}}(\bvec{k}) = i \delta_{\bm{\alpha}\mathbf{t}} \delta^{\rho+} \frac{k_t}{|\bvec{k}|} \gamma(\bvec{k}). \label{eq: source gamma}
\end{align}

With the quadratic operator $\mathbb{M}_{\bm{\alpha\beta}}^{\rho\sigma}(\bvec{k})$ and the source term $j_{\bm{\alpha}}^\rho(\bvec{k})$ defined, the square in the Gaussian action can be completed by shifting the boundary fields as ${{b'}_{\bm{\alpha}}^\rho = b_{\bm{\alpha}}^\rho + \frac{1}{2} \left(\mathbb{M}(\bvec{k})^{-1}\right)_{\bm{\alpha\beta}}^{\rho\sigma}} \, j_{\bm{\beta}}^\sigma(\bvec{k})$, reducing the integrand to
\begin{align*}
	& b_{\bm{\alpha}}^\rho(-\bvec{k}) \mathbb{M}^{\rho\sigma}_{\bm{\alpha\beta}}(\bvec{k}) b_\mathbf{\bm{\beta}}^\sigma(\bvec{k}) + b_{\bm{\alpha}}^\rho(-\bvec{k}) j_{\bm{\alpha}}^\rho(\bvec{k})\\
	& = {b'}_{\bm{\alpha}}^\rho(-\bvec{k}) \mathbb{M}^{\rho\sigma}_{\bm{\alpha\beta}}(\bvec{k}) {b'}_{\bm{\beta}}^\sigma(\bvec{k}) - \frac{1}{2} j_{\bm{\alpha}}^\rho(-\bvec{k}) {b'}_{\bm{\alpha}}^\rho(\bvec{k}) + \frac{1}{2} {b'}_{\bm{\alpha}}^\rho(-\bvec{k}) j_{\bm{\alpha}}^\rho(\bvec{k}) - \frac{1}{4} j_{\bm{\alpha}}^\rho(-\bvec{k}) \left(\mathbb{M}(\bvec{k})^{-1}\right)_{\bm{\alpha\beta}}^{\rho\sigma} \, j_{\bm{\beta}}^\sigma(\bvec{k}).
\end{align*}
Given that the quadratic operator satisfies the symmetry property $\mathbb{M}_{\bm{\alpha\beta}}^{\rho\sigma}(\bvec{k}) = \mathbb{M}_{\bm{\beta\alpha}}^{\sigma\rho}(-\bvec{k})$, it follows that its inverse inherits the same symmetry, $\left(\mathbb{M}(\bvec{k})^{-1}\right)_{\bm{\alpha\beta}}^{\rho\sigma} \, = \left(\mathbb{M}(-\bvec{k})^{-1}\right)_{\bm{\beta\alpha}}^{\sigma\rho}$, a relation that was used in the derivation above. Under the momentum integral in the action the second and third terms will cancel each other, effectively decoupling the field $b_{\bm{\alpha}}^\rho$ from $\gamma$ in the action. This implies that we can factorize the partition function as
\begin{align*}
	Z^{(L)}_{b^+, b^-, \gamma} & = Z^{(L)}_{b^+, b^-} Z^{(L)}_{\gamma}.
\end{align*}
Both partition functions are now Gaussian functional integrals that can be performed analytically. For the first partition function this gives
\begin{align*}
	Z^{(L)}_{b^+, b^-} & = \int \DD{{b'}^+} \DD{{b'}^-} \exp \left[-\frac{1}{2} \int \! \frac{\mathrm{d}^3\bvec{k}}{(2\pi)^3} \left\{{b'}_{\bm{\alpha}}^\rho(-\bvec{k}) \mathbb{M}^{\rho\sigma}_{\bm{\alpha\beta}}(\bvec{k}) {b'}_\mathbf{\bm{\beta}}^\sigma(\bvec{k})\right\}\right] = \frac{C_b}{\sqrt{\det{\mathbb{M}}}}.
\end{align*}
with $C_b$ an infinite multiplicative factor independent of the interplate distance $L$. Filling in the expression for the source field $j_{\bm{\alpha}}^\rho(\bvec{k})$ in terms of $\gamma$, equation \eqref{eq: source gamma}, we can also perform the functional integral over $\gamma$ in $Z^{(L)}_{\gamma}$ which results in
\begin{align}
	Z^{(L)}_{\gamma} & = \int \DD{\gamma} \exp \left[-\frac{1}{2} \int \! \frac{\mathrm{d}^3\bvec{k}}{(2\pi)^3} \left\{j_{\bm{\alpha}}^\rho(-\bvec{k}) \left[-\frac{1}{4}\left(\mathbb{M}(\bvec{k})^{-1}\right)_{\bm{\alpha\beta}}^{\rho\sigma}\right] \, j_{\bm{\beta}}^\sigma(\bvec{k})\right\}\right] \nonumber\\
	& = \int \DD{\gamma} \exp \left[-\frac{1}{2} \int \! \frac{\mathrm{d}^3\bvec{k}}{(2\pi)^3} \left\{\gamma(-\bvec{k}) \left[-\frac{k_t^2}{4\bvec{k}^2} \left(\mathbb{M}(\bvec{k})^{-1}\right)_{\mathbf{tt}}^{++} \right] \gamma(\bvec{k}) \right\}\right] \nonumber \\
	& = \int \DD{\gamma} \exp \left[-\frac{1}{2} \int \! \frac{\mathrm{d}^3\bvec{k}}{(2\pi)^3} \left\{\gamma(-\bvec{k}) \left(\mathbb{M}(\bvec{k})^{-1}\right)_{\mathbf{tt}}^{++} \mathbb{O}(\bvec{k}) \gamma(\bvec{k}) \right\}\right] \nonumber\\
	& = \frac{C_{\gamma}}{\sqrt{\det\left(\left(\mathbb{M}^{-1}\right)_{\mathbf{tt}}^{++} \cdot \mathbb{O}\right)}}, \label{eq: part function gamma Landau}
\end{align}
in which $C_\gamma$ is an infinite multiplicative constant independent of the interplate distance $L$ and the $L$-independent operator $\mathbb{O}(\bvec{k})$ was defined as
\begin{align*}
	\mathbb{O}(\bvec{k}) & = -\frac{k_t^2}{4\bvec{k}^2}.
\end{align*}
The zero modes of this operator are of no immediate concern, since these will anyhow not affect the physically $L$-dependent contributions we are after. Furthermore, the $\det$ can be regularized in whatever fashion, it will drop out since it appears in a $\det \mathbb{O}/\det \mathbb{O}$ combination, see below.

With the functional integrals evaluated, we are now able to calculate the Casimir energy using equation \eqref{eq: Casimir energy split}.
If we expand the functional determinant in its Fourier modes, i.e.~$\det M = \exp\left[\sum_{k} \ln \left|M_k\right|\right]$
with $\left|M_k\right|$ matrix determinants, and write $\mathbb{M} \xrightarrow{L \rightarrow \infty} \tilde{\mathbb{M}}$, we find
\begin{align*}
	& E_{\text{Cas}}(L) = -\lim_{T \rightarrow \infty} \frac{1}{T}\ln\left[\frac{Z^{(L)}_{A, b^+, b^-, h, \gamma}}{Z^{(\infty)}_{A, b^+, b^-, h, \gamma}}\right] = -\lim_{T \rightarrow \infty} \frac{1}{T}\ln\left[\frac{Z^{(L)}_{b^+, b^-} \, Z^{(L)}_{\gamma}}{Z^{(\infty)}_{b^+, b^-} \, Z^{(\infty)}_{\gamma}}\right]\\
	& = -\lim_{T \rightarrow \infty} \frac{1}{T}\ln\left[\sqrt{\frac{\det \tilde{\mathbb{M}} \cdot \det((\tilde{\mathbb{M}}^{-1})^{++}_{\mathbf{tt}}) \cdot \det \mathbb{O}}{\det \mathbb{M} \cdot \det((\mathbb{M}^{-1})^{++}_{\mathbf{tt}}) \cdot \det\mathbb{O}}}\right] = \lim_{T \rightarrow \infty} \frac{1}{2T}\ln\left[\frac{\det ( \mathbb{M} \cdot (\mathbb{M}^{-1})^{++}_{\mathbf{tt}})}{\det ( \tilde{\mathbb{M}} \cdot (\tilde{\mathbb{M}}^{-1})^{++}_{\mathbf{tt}})}\right]\\
	& = \lim_{T \rightarrow \infty} \frac{1}{2T} T \ell_x \ell_y \int \! \frac{\mathrm{d}^3 \bvec{k}}{(2\pi)^3} \ln \frac{|\mathbb{M} \cdot (\mathbb{M}^{-1})^{++}_{\mathbf{tt}} |}{|\tilde{\mathbb{M}} \cdot (\tilde{\mathbb{M}}^{-1})^{++}_{\mathbf{tt}}|}
	= \frac{\ell_x \ell_y}{2} \int \! \frac{\mathrm{d}^3 \bvec{k}}{(2\pi)^3} \ln \frac{|\mathbb{M}| \, |(\mathbb{M}^{-1})^{++}_{\mathbf{tt}} |}{|\tilde{\mathbb{M}}| \, |(\tilde{\mathbb{M}}^{-1})^{++}_{\mathbf{tt}}|}
\end{align*}
where $\ell_x, \ell_y$ and $T$ denote the infinite spacetime extents in the $x$-, $y$- and $t$-direction respectively. Focusing on the determinants in the numerator, using the block matrices defined in \eqref{eq: quadratic operator block matrix structure Landau} and Appendix \ref{ch: Schur}, which introduces the Schur complement $\mathbb{M}/\mathbf{A}$, we obtain by use of equation \eqref{eq: Schur determinant} that
\begin{align}
	|\mathbb{M}| \, |(\mathbb{M}^{-1})^{++}_{\mathbf{tt}} | = |\mathbf{A}| \cdot |\mathbb{M}/\mathbf{A}| \cdot |(\mathbb{M}/\mathbf{A})^{-1}| = |\mathbf{A}|
	\label{eq: Schur cancellation}
\end{align}
since $(\mathbb{M}^{-1})^{++}_{\mathbf{tt}} = (\mathbb{M}/\mathbf{A})^{-1}$, by virtue of expression \eqref{eq: Schur inverse}. Computing the determinant of this $5\times5$ submatrix $\mathbf{A}$ after simplification gives
\begin{align*}
	|\mathbf{A}| = -\lambda \frac{k_t^2 \bvec{k}^2}{16} \left(1-e^{-2i\theta}e^{-2|\bvec{k}|L}\right) \cdot \left(1-e^{2i\theta}e^{-2|\bvec{k}|L}\right).
\end{align*}
To conclude, it follows that the Casimir energy in the Landau gauge is given by
\begin{align}
	E_{\text{Cas}}(L,\theta) & = \frac{\ell_x \ell_y}{2} \int \! \frac{\mathrm{d}^3 \bvec{k}}{(2\pi)^3} \ln \frac{|\mathbf{A}|}{|\tilde{\mathbf{A}}|} = \frac{\ell_x \ell_y}{2} \int \! \frac{\mathrm{d}^3 \bvec{k}}{(2\pi)^3} \ln\left[\left(1-e^{-2i\theta}e^{-2|\bvec{k}|L}\right) \left(1-e^{2i\theta}e^{-2|\bvec{k}|L}\right)\right] \nonumber\\
	& = \frac{1}{4\pi^2} \frac{\ell_x \ell_y}{(2L)^3} \left\{\int_0^{+\infty} \mathrm{d}x \ x^2 \ln(1-e^{-2i\theta}e^{-x}) + \int_0^{+\infty} \mathrm{d}x \ x^2 \ln(1-e^{2i\theta}e^{-x})\right\} \nonumber\\
	& = -\frac{\ell_x \ell_y}{8 \pi^2 L^3} \mathrm{Re} \, \mathrm{Li}_4(e^{2i\theta}), \label{eq: Casimir energy non-ghost Landau}
\end{align}
in which we reused the notation with tilde to denote the limit of infinitely separated plates i.e., $\mathbf{A} \xrightarrow{L \rightarrow \infty} \tilde{\mathbf{A}}$. Here $\mathrm{Re} \, \mathrm{Li}_4(...)$ denotes the real part of the polylogarithm of order four.

As a final remark on the computation in the Landau gauge, let us take a closer look at equation \eqref{eq: Schur cancellation}. We see that by using the Schur complement $\mathbb{M}/\mathbf{A}$, we were able to rewrite the determinant of the quadratic operator $\mathbb{M}$ such that the contributions from the field $b^+_t$ cancel, leaving no contribution to the Casimir energy. Remembering that this field is the edge mode living on the {DEM} plate, we conclude that the edge mode does not contribute to the Casimir energy and force. This result is consistent with the Casimir energy calculation in the {DEM}--{DEM} setup, where the contributions of the edge modes on both plates also cancel, rendering their contribution to the Casimir energy zero \cite{canforaDynamicalEdgeModes2025}.

\subsection{Integrating out edge fields in Coulomb gauge}\label{sssec: edge fields PEMC - DEM Coulomb}
We now repeat the above computation but in Coulomb gauge $\mathcal{F}_{\mathtt{C}}[A] = \partial_m A_m$, in which the partition function is given by
\begin{align}
	Z^{(L)}_{b^+, b^-, \gamma} = \int \DD{b^+}\DD{b^-}\DD{\gamma} \ \exp\left[-\frac{1}{2} \int \! \frac{\mathrm{d}^4k}{(2\pi)^4} \, \left\{ v^{\mathtt{C}}_\mu(k)(\mathbb{K}^{\mathtt{C}})^{-1}_{\mu\nu}v^{\mathtt{C}}_\nu(-k)\right\}\right],
	\label{eq: part function coulomb PEMC-DEM}
\end{align}
where the inverse of the quadratic operator in the Coulomb gauge takes the form
\begin{align*}
	(\mathbb{K}^{\mathtt{C}})^{-1}_{\mu\nu} = \frac{1}{k^2}\left[\xi\frac{k^4}{\vec{k}^4}L_{\mu\nu}(k) + \delta_{\mu\nu} + \frac{k_t^2}{\vec{k}^2}\delta_{t\mu}\delta_{t\nu} - \frac{1}{\vec{k}^2} k_m k_n\delta_{m\mu}\delta_{n\nu}\right].
\end{align*}
Writing the field $v^{\mathtt{C}}_\mu(\bvec{k})$ as
\begin{align*}
	v^{\mathtt{C}}_\mu(\bvec{k}) = \left[w_\mu(\bvec{k}) + m_\mu(\bvec{k})\right] + u^{\mathtt{C}}_\mu(\bvec{k})
\end{align*}
with $w_\mu(\bvec{k}), m_\mu(\bvec{k})$ and $u^{\mathtt{C}}_\mu(\bvec{k})$ defined by equations \eqref{eq: PEMC source}, \eqref{eq: DEM source} and \eqref{eq: gamma coulomb source} respectively, the integrand in \eqref{eq: part function coulomb PEMC-DEM} can be expanded, giving
\begin{align}
	\begin{split}
		& v^{\mathtt{C}}_\mu(k) (\mathbb{K}^{\mathtt{C}})^{-1}_{\mu\nu} v^{\mathtt{C}}_\nu(-k)\\
		& = \left[w_\mu(k) + m_\mu(k)\right] (\mathbb{K}^{\mathtt{C}})^{-1}_{\mu\nu} \left[w_\nu(-k) + m_\nu(-k)\right] + u^{\mathtt{C}}_\mu(k) (\mathbb{K}^{C})^{-1}_{\mu\nu} u^{\mathtt{C}}_\nu(-k)\\
		& \quad + \left[w_\mu(k) + m_\mu(k)\right] (\mathbb{K}^{\mathtt{C}})^{-1}_{\mu\nu} u^{\mathtt{C}}_\nu(-k) + u^{\mathtt{C}}_\mu(k) (\mathbb{K}^{\mathtt{C}})^{-1}_{\mu\nu} \left[w_\nu(-k) + m_\nu(-k)\right].
	\end{split}
	\label{eq: expand kmunuC}
\end{align}

Focusing on the last three terms first, given that
\begin{align*}
	u^{\mathtt{C}}_\mu(k) L_{\mu\nu}(k) & = -\frac{i}{\xi} \delta_{i\mu} k_i \gamma(\bvec{k}) e^{i k_z z^+} \frac{k_\mu k_\nu}{k^2} = -\frac{i}{\xi} \frac{\vec{k}^2}{k^2} k_\nu \gamma(\bvec{k}) e^{i k_z z^+}\\
	u^{\mathtt{C}}_\mu(k) \frac{k_t^2}{\vec{k}^2} \delta_{t\mu} \delta_{t\nu} & = 0\\
	u^{\mathtt{C}}_\mu(k) \frac{1}{\vec{k}^2} k_i k_j \delta_{i\mu} \delta_{j\nu} & = -\frac{i}{\xi} \delta_{l\mu} k_l \gamma(\bvec{k}) e^{i k_z z^+} \frac{1}{\vec{k}^2} k_i k_j \delta_{i\mu} \delta_{j\nu} = -\frac{i}{\xi} \delta_{j\nu} k_j \gamma(\bvec{k}) e^{i k_z z^+} = u^{\mathtt{C}}_\nu(k)
\end{align*}
one finds that
\begin{align*}
	u^{\mathtt{C}}_\mu(k) (\mathbb{K}^{\mathtt{C}})^{-1}_{\mu\nu} & = \frac{1}{k^2}\left[-i \frac{k^2}{\vec{k}^2} k_\nu \gamma(\bvec{k}) e^{i k_z z^+} + u^{\mathtt{C}}_\nu(k) + 0 - u^{\mathtt{C}}_\nu(k)\right] = -\frac{i}{\vec{k}^2} k_\nu \gamma(\bvec{k}) e^{i k_z z^+}.
\end{align*}
Hence for the second term in \eqref{eq: expand kmunuC}, we find the term quadratic in the field $\gamma$,
\begin{align*}
	u^{\mathtt{C}}_\mu(k) (\mathbb{K}^{\mathtt{C}})^{-1}_{\mu\nu} u^{\mathtt{C}}_\nu(-k) & = -\frac{i}{\vec{k}^2} k_\nu \gamma(\bvec{k}) e^{i k_z z^+} u^{\mathtt{C}}_\nu(-k) = -\frac{i}{\vec{k}^2} k_\nu \gamma(\bvec{k}) e^{i k_z z^+} \frac{i}{\xi} \delta_{j\nu} k_j \gamma(-\bvec{k}) e^{-i k_z z^+} = \frac{1}{\xi} \gamma(\bvec{k}) \gamma(\bvec{k}),
\end{align*}
which in the action will give the same contribution as equation \eqref{eq: Sgammaquad} in the Landau gauge, and will therefore also vanish under dimensional regularisation.

From
\begin{align}
	\!\!
	\begin{split}
		k_\mu m_\mu(-k) & = k_t m_t(-k) + k_a m_a(-k) + k_z m_z(-k)\\
		& = \left[k_t b_t^+(-\bvec{k}) + i k_a k_z b_a^+(-\bvec{k}) - i k_z k_a b_a^+(-\bvec{k})\right] e^{-i k_z z^+} = k_t b_t^+(-\bvec{k}) e^{-i k_z z^+},
	\end{split}
	\! \label{eq: }
\end{align}
and $k_\nu w_\nu(-k) = 0$, it follows that the last term in \eqref{eq: expand kmunuC} reduces to
\begin{align*}
	& u^{\mathtt{C}}_\mu(k) (\mathbb{K}^{\mathtt{C}})^{-1}_{\mu\nu} \left[w_\nu(-k) + m_\nu(-k)\right] = -\frac{i}{\vec{k}^2} k_\nu \gamma(\bvec{k}) e^{i k_z z^+} \left[w_\nu(-k) + m_\nu(-k)\right]\\
	& = -\frac{i}{\vec{k}^2} \gamma(\bvec{k}) e^{i k_z z^+} k_\nu m_\nu(-k) = -i\frac{k_t}{\vec{k}^2} b_t^+(-\bvec{k}) \gamma(\bvec{k})
\end{align*}
and similarly for the third term
\begin{align*}
	\left[w_\mu(k) + m_\mu(k)\right] (\mathbb{K}^{\mathtt{C}})^{-1}_{\mu\nu} u^{\mathtt{C}}_\nu(-k) & = i\frac{k_t}{\vec{k}^2} b_t^+(\bvec{k}) \gamma(-\bvec{k}).
\end{align*}
Let us now switch focus to the terms quadratic in the field $b_A^\rho(\bvec{k})$, given by \linebreak $\left[w_\mu(k) + m_\mu(k)\right] (\mathbb{K}^{\mathtt{C}})^{-1}_{\mu\nu} \left[w_\nu(-k) + m_\nu(-k)\right]$. When one acts with the inverse quadratic operator $(\mathbb{K}^{\mathtt{C}})^{-1}_{\mu\nu}$ on $w_\mu(k)$, it can be found that
\begin{align*}
	w_\mu(k) (\mathbb{K}^{\mathtt{C}})^{-1}_{\mu\nu} & = \frac{1}{k^2} \left[w_\nu(k) + \frac{1}{\vec{k}^2} \Big(k_t w_t(k) k_t \delta_{t\nu} - k_i w_i(k) k_j \delta_{j\nu}\Big)\right]\\
	& = \frac{1}{k^2} \left[w_\nu(k) + \frac{k_t w_t(k)}{\vec{k}^2} \Big(k_t \delta_{t\nu} + k_j \delta_{j\nu}\Big)\right] = \frac{1}{k^2} \left[w_\nu(k) + \frac{k_t w_t(k)}{\vec{k}^2} k_\nu \right]
\end{align*}
where in the first and second line we used that $0 = k_\nu w_\mu(k) \Leftrightarrow k_t w_t(k) = -k_i w_i(k)$. Consequently, this implies that
\begin{align*}
	& w_\mu(k) (\mathbb{K}^{\mathtt{C}})^{-1}_{\mu\nu} w_\nu(-k) = \frac{1}{k^2} w_\nu(k) w_\nu(-k),
\end{align*}
and also
\begin{align*}
	& w_\mu(k) (\mathbb{K}^{\mathtt{C}})^{-1}_{\mu\nu} m_\nu(-k) = \frac{1}{k^2} w_\nu(k) m_\nu(-k) + \frac{k_t w_t(k)}{k^2 \vec{k}^2} k_\nu m_\nu(-k)\\
	& = \frac{1}{k^2} w_\nu(k) m_\nu(-k) + \frac{k_t^2}{k^2 \vec{k}^2} w_t(k) b_t^+(-\bvec{k}) e^{-i k_z z^+}\\
	& = \frac{1}{k^2} w_\nu(k) m_\nu(-k) + \frac{k_t^2}{k^2 \vec{k}^2} b_t^+(-\bvec{k}) \left[i \cos(\theta) k_z b^-_t(\bvec{k}) - \varepsilon_{34BC} \sin(\theta) k_B b^-_C(\bvec{k})\right] e^{-i k_z L}
\end{align*}
and similarly by replacing $k_\mu$ with $-k_\mu$,
\begin{align*}
	& m_\mu(k) (\mathbb{K}^{\mathtt{C}})^{-1}_{\mu\nu} w_\nu(-k)\\
	& = \frac{1}{k^2} m_\nu(k) w_\nu(-k) - \frac{k_t^2}{k^2 \vec{k}^2} b_t^+(\bvec{k}) \left[i \cos(\theta) k_z b^-_t(-\bvec{k}) - \varepsilon_{34BC} \sin(\theta) k_B b^-_C(-\bvec{k})\right] e^{i k_z L}.
\end{align*}

\begin{samepage}
\noindent Finally we need to consider the term $m_\mu(k) (\mathbb{K}^{\mathtt{C}})^{-1}_{\mu\nu} m_\nu(-k)$, which yields
\begin{align*}
	& m_\mu(k) (\mathbb{K}^{\mathtt{C}})^{-1}_{\mu\nu} m_\nu(-k)\\
	& = \frac{1}{k^2}\left[\xi\frac{k^4}{\vec{k}^4} m_\mu(k) L_{\mu\nu}(k) m_\nu(-k) + m_\nu(k) m_\nu(-k) + \frac{k_t^2}{\vec{k}^2} m_t(k) m_t(-k) - \frac{1}{\vec{k}^2} k_i m_i(k) k_j m_j(-k)\right]\\
	& = \frac{1}{k^2}\left[\xi\frac{k^4}{\vec{k}^4} m_\mu(k) L_{\mu\nu}(k) m_\nu(-k) + m_\nu(k) m_\nu(-k) + b_t^+(-\bvec{k}) \frac{k_t^2}{\vec{k}^2} b_t^+(\bvec{k}) \right]\\
	& = \frac{1}{k^2}\left[\frac{k^2}{\vec{k}^2} \left(\xi\frac{k^2}{\vec{k}^2} + 1\right) m_\mu(k) L_{\mu\nu}(k) m_\nu(-k) + m_\nu(k) m_\nu(-k)\right]\\
	& = \frac{1}{\vec{k}^2} \left(\xi\frac{k^2}{\vec{k}^2} + 1\right) m_\mu(k) L_{\mu\nu}(k) m_\nu(-k) + \frac{1}{k^2} m_\nu(k) m_\nu(-k),
\end{align*}
where we used that $m_\mu(k) L_{\mu\nu} m_\nu(-k) = b_t^+(-\bvec{k}) \dfrac{k_t^2}{k^2} b_t^+(\bvec{k})$.
\end{samepage}
\enlargethispage{\baselineskip}

Therefore, combining the results above we find for the terms in the action quadratic in the field $b_A^\rho(\bvec{k})$ that
\begin{align*}
	& \left[w_\mu(k) + m_\mu(k)\right] (\mathbb{K}^{\mathtt{C}})^{-1}_{\mu\nu} \left[w_\nu(-k) + m_\nu(-k)\right]\\
	& = \frac{1}{k^2} \left[w_\mu(k) w_\mu(-k) + w_\mu(k) m_\mu(-k) + m_\mu(k) w_\mu(-k) + m_\mu(k) m_\mu(-k) \right]\\
	& \quad {+ \frac{1}{\vec{k}^2} \left(\xi\frac{k^2}{\vec{k}^2} + 1\right) m_\mu(k) L_{\mu\nu}(k) m_\nu(-k)}\\
	& \quad {+ \frac{k_t^2}{k^2 \vec{k}^2} b_t^+(-\bvec{k}) \left[i \cos(\theta) k_z b^-_t(\bvec{k}) - \varepsilon_{34BC} \sin(\theta) k_B b^-_C(\bvec{k})\right] e^{-i k_z L}}\\
	& \quad {- \frac{k_t^2}{k^2 \vec{k}^2} b_t^+(\bvec{k}) \left[i \cos(\theta) k_z b^-_t(-\bvec{k}) - \varepsilon_{34BC} \sin(\theta) k_B b^-_C(-\bvec{k})\right] e^{i k_z L}}.
\end{align*}
Since the definitions of the fields $w_\mu(\bvec{k})$ and $m_\mu(\bvec{k})$ do not depend on the gauge choice (see equations \eqref{eq: PEMC source}, \eqref{eq: DEM source}), we can compare the above with the expanded integrand in the Landau gauge, see equation \eqref{eq: integrand full PEMC-DEM}. It can be observed that the first line is the same as in the Landau gauge, while the second line term is also present in the Landau gauge but with prefactor $\frac{\xi-1}{k^2}$ instead of $\frac{1}{\vec{k}^2} \left(\xi\frac{k^2}{\vec{k}^2} + 1\right)$ here in the Coulomb gauge. Compared with equation \eqref{eq: integrand full PEMC-DEM}, the last two lines of terms are new.

As in the Landau gauge, the partition function for plates separated by a distance $L$ thus takes the form
\begin{align*}
	Z^{(L)}_{b^+, b^-, \gamma} = \! \int \! \mathcal{D} b^+ \, \mathcal{D} b^- \, \mathcal{D}\gamma \, \exp \left[-\frac{1}{2} \int \! \frac{\mathrm{d}^3\bvec{k}}{(2\pi)^3} \left\{b_A^\rho(-\bvec{k}) \mathbb{C}^{\rho\sigma}_{AB}(\bvec{k}) b_B^\sigma(\bvec{k}) + b_t^+(-\bvec{k}) j(\bvec{k})\right\}\right]
\end{align*}
with the important difference that the quadratic operator $\mathbb{C}^{\rho\sigma}_{AB}(\bvec{k})$ in the Coulomb gauge differs from the quadratic operator in the Landau gauge $\mathbb{M}^{\rho\sigma}_{AB}(\bvec{k})$ as
\begin{equation}
\begin{minipage}{0.9\textwidth}
\centering
$\begin{aligned}
	\mathbb{C}^{++}_{tt}(\bvec{k}) & = -\frac{1}{4|\bar{k}|^3} \left(\xi k_t^2 + 2 |\bar{k}|^2 \right), & \mathbb{C}^{+-}_{tt}(\bvec{k}) & = -\frac{\cos(\theta)}{2} e^{-|\bar{k}|L}, & \mathbb{C}^{-+}_{tt}(\bvec{k}) & = -\frac{\cos(\theta)}{2} e^{-|\bar{k}|L}
\end{aligned}$\\
$\begin{aligned}
	\mathbb{C}^{+-}_{tb}(\bvec{k}) & = \varepsilon_{34ab} k_a \sin(\theta) \frac{e^{-|\bar{k}|L}}{2|\bar{k}|}, & \mathbb{C}^{-+}_{bt}(\bvec{k}) & = -\varepsilon_{34ab} k_a \sin(\theta) \frac{e^{-|\bar{k}|L}}{2|\bar{k}|},
\end{aligned}$
\end{minipage}
\label{eq: matrix C coulomb}
\end{equation}
with the notation $|\bar{k}|$ defined as $|\bar{k}| = \sqrt{k_x^2+k_y^2}$. Looking at the other components of the quadratic operator $\mathbb{C}$, it can be seen that these do not differ from the quadratic operator in the Landau gauge i.e., $\mathbb{C}^{\rho\sigma}_{AB}(\bvec{k}) = \mathbb{M}^{\rho\sigma}_{AB}(\bvec{k})$. Comparing the action in the Landau and the Coulomb gauge, it can be observed that also the source field $s_{\mathtt{C}}(\bvec{k})$ is different in the Coulomb gauge.

We now can change the basis to $(\textbf{e}_{\mathbf{1}}, \textbf{e}_{\mathbf{2}}, \textbf{e}_{\mathbf{t}})$ and complete the square such that the partition function factorises as
\begin{align*}
	Z^{(L)}_{b^+, b^-, \gamma} & = Z^{(L)}_{b^+, b^-} Z^{(L)}_{\gamma}.
\end{align*}
with
\begin{align*}
	Z^{(L)}_{b^+, b^-} & = \int \mathcal{D} {b'}^+ \, \mathcal{D} {b'}^- \, \exp \left[-\frac{1}{2} \int \! \frac{\mathrm{d}^3\bvec{k}}{(2\pi)^3} \left\{{b'}_{\bm{\alpha}}^\rho(-\bvec{k}) \mathbb{C}^{\rho\sigma}_{\bm{\alpha\beta}}(\bvec{k}) {b'}_\mathbf{\bm{\beta}}^\sigma(\bvec{k})\right\}\right] = \frac{C_b}{\sqrt{\det{\mathbb{C}}}}.
\end{align*}
and
\begin{align*}
	Z^{(L)}_{\gamma} & = \int \mathcal{D} \gamma \, \exp \left[-\frac{1}{2} \int \! \frac{\mathrm{d}^3\bvec{k}}{(2\pi)^3} \left\{j_{\bm{\alpha}}^\rho(-\bvec{k}) \left[-\frac{1}{4}\left(\mathbb{M}(\bvec{k})^{-1}\right)_{\bm{\alpha\beta}}^{\rho\sigma}\right] \, j_{\bm{\beta}}^\sigma(\bvec{k})\right\}\right] = \frac{C_{\gamma}}{\sqrt{\det\left(\left(\mathbb{C}^{-1}\right)_{\mathbf{tt}}^{++} \cdot \mathbb{O}\right)}}.
\end{align*}
Here the source field $j_{\bm{\alpha}}^\rho(\bvec{k})$ in the Coulomb gauge is defined as
\begin{align*}
	j_{\bm{\alpha}}^\rho(\bvec{k}) = \delta_{\bm{\alpha}\mathbf{t}} \delta^{\rho+} s_{\mathtt{C}}(\bvec{k}) = i \delta_{\bm{\alpha}\mathbf{t}} \delta^{\rho+} \frac{k_t}{|\bar{k}|} \gamma(\bvec{k}),
\end{align*}
and consequently the quadratic operator $\mathbb{O}$ in the Coulomb gauge is simply the same $L$-independent operator as in the Landau gauge, with the exception of the replacement $|\bvec{k}| \rightarrow |\bar{k}|$
\begin{align*}
	\mathbb{O} & = -\frac{k_t^2}{4\bar{k}^2}.
\end{align*}

Finally, focusing on the Casimir energy calculation, we see that we will need the equivalent of the $5 \times 5$ submatrix $\mathbf{A}$ of $\mathbb{M}$ for the matrix $\mathbb{C}$. Remembering how the quadratic operator in the Coulomb gauge $\mathbb{C}^{\rho\sigma}_{AB}(\bvec{k})$ differs from $\mathbb{M}^{\rho\sigma}_{AB}(\bvec{k})$ in the Landau gauge, see \eqref{eq: matrix C coulomb}, it can be observed that the only differences are in the components with $(\rho,A) = (+,t)$ or $(\sigma,B) = (+,t)$. Put differently, neglecting the difference in the sources for a moment, the actions in the Landau and Coulomb gauge differ in the terms with $b_t^+(\bvec{k})$ or $b_t^+(-\bvec{k})$. This has as consequence that in the Coulomb gauge, only the submatrices $\mathbf{B}, \mathbf{C}$ and $\mathbf{D}$ are modified relative to the Landau gauge. The $5 \times 5$ submatrix $\mathbf{A}$ of the Coulomb quadratic operator $\mathbb{C}$ is precisely the same as the submatrix $\mathbf{A}$ for the Landau quadratic operator $\mathbb{M}$, see also matrix \eqref{eq: quadratic operator block matrix structure coulomb}, below

\renewcommand{\arraystretch}{1.4}
\begin{align}
	\mathbb{C}_{\bm{\alpha\beta}}^{\rho\sigma}(\bvec{k}) = \left(
	\begin{array}{ccccc|c}
		\mathbb{M}_{\mathbf{11}}^{--} & \mathbb{M}_{\mathbf{11}}^{-+} & \mathbb{M}_{\mathbf{12}}^{--} & \mathbb{M}_{\mathbf{12}}^{-+} & \mathbb{M}_{\mathbf{1t}}^{--} & \mathbb{M}_{\mathbf{1t}}^{-+}\\
		\mathbb{M}_{\mathbf{11}}^{+-} & \mathbb{M}_{\mathbf{11}}^{++} & \mathbb{M}_{\mathbf{12}}^{+-} & \mathbb{M}_{\mathbf{12}}^{++} & \mathbb{M}_{\mathbf{1t}}^{+-} & \mathbb{M}_{\mathbf{1t}}^{++}\\
		\mathbb{M}_{\mathbf{21}}^{--} & \mathbb{M}_{\mathbf{21}}^{-+} & \mathbb{M}_{\mathbf{22}}^{--} & \mathbb{M}_{\mathbf{22}}^{-+} & \mathbb{M}_{\mathbf{2t}}^{--} & -\frac{e^{-|\bar{k}|L} \sin(\theta)}{2}\\
		\mathbb{M}_{\mathbf{21}}^{+-} & \mathbb{M}_{\mathbf{21}}^{++} & \mathbb{M}_{\mathbf{22}}^{+-} & \mathbb{M}_{\mathbf{22}}^{++} & \mathbb{M}_{\mathbf{2t}}^{+-} & \mathbb{M}_{\mathbf{2t}}^{++}\\
		\mathbb{M}_{\mathbf{t1}}^{--} & \mathbb{M}_{\mathbf{t1}}^{-+} & \mathbb{M}_{\mathbf{t2}}^{--} & \mathbb{M}_{\mathbf{t2}}^{-+} & \mathbb{M}_{\mathbf{tt}}^{--} & -\frac{e^{-|\bar{k}|L} \cos(\theta)}{2}\\ \hline
		\mathbb{M}_{\mathbf{t1}}^{+-} & \mathbb{M}_{\mathbf{t1}}^{++} & -\frac{e^{-|\bar{k}|L} \sin(\theta)}{2} & \mathbb{M}_{\mathbf{t2}}^{++} & -\frac{e^{-|\bar{k}|L} \cos(\theta)}{2} & -\frac{\xi k_t^2 + 2 \bar{k}^2}{4|\bar{k}|^3}
	\end{array}
	\right).
	\label{eq: quadratic operator block matrix structure coulomb}
\end{align}

Consequently we retrieve the same contribution to the Casimir energy from the non-ghost fields in the Coulomb gauge, namely
\begin{align}
	\begin{split}
		E_{\text{non-ghost}}(L) & = \frac{l_x l_y}{2} \int \! \frac{\mathrm{d}^3 \bvec{k}}{(2\pi)^3} \ln \frac{|\mathbb{C}| \, |(\mathbb{C}^{-1})^{++}_{\mathbf{tt}} |}{|\tilde{\mathbb{C}}| \, |(\tilde{\mathbb{C}}^{-1})^{++}_{\mathbf{tt}}|} =
		\frac{l_x l_y}{2} \int \! \frac{\mathrm{d}^3 \bvec{k}}{(2\pi)^3} \ln \frac{|\mathbf{A}|}{|\tilde{\mathbf{A}}|}\\
		& = -\frac{l_x l_y}{8 \pi^2 L^3} \mathrm{Re} \, \mathrm{Li}_4(e^{2i\theta}).
	\end{split}
	\label{eq: Casimir energy non-ghost Coulomb}
\end{align}

\subsection{Casimir energy and force}\label{sec:casimir}
Because of translational symmetry along the plates, i.e.~the $x$- and $y$-directions, the Casimir energy per unit area for the {PEMC}--{DEM} case can immediately be read off from Eq.~\eqref{eq: Casimir energy non-ghost Landau}:
\begin{align*}
	\mathcal{E}_{\text{Cas}}(L,\theta) = -\frac{1}{8 \pi^2 L^3} \mathrm{Re} \, \mathrm{Li}_4(e^{2i\theta}).
\end{align*}
Finally, taking the derivative with respect to the plate separation $L$, we retrieve the Casimir force per unit area
\begin{align}
	\mathcal{F}_{\text{Cas}}(L,\theta) = -\frac{\partial \mathcal{E}_{\text{Cas}}}{\partial L} = -\frac{3}{8 \pi^2 L^4} \mathrm{Re} \, \mathrm{Li}_4(e^{2i\theta}). \label{eq: Casimir force PEMC-DEM}
\end{align}

\begin{figure}[h]
	\centering
	\includegraphics[width=0.75\textwidth]{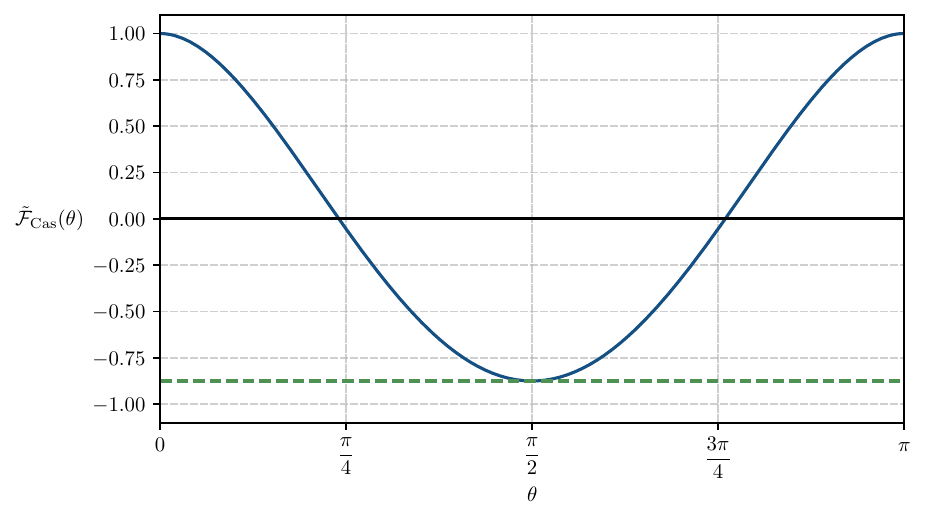}
	\caption{The relative Casimir force $\tilde{\mathcal{F}}_{\text{Cas}}(\theta)$ as a function of the duality angle \(\theta\) of the {PEMC} plate, cfr.~Eq.~\eqref{eq:F-tilde}.}
	\label{fig: Relative Casimir PEMC-DEM}
\end{figure}

A first, rather simple observation that we could make is that the Casimir force depends on the duality angle of the {PEMC} plate. For instance, the force between the plates in a {PMC}--{DEM} setup ($\theta = 0$) reduces to the same attractive force between two identical {PEMC} plates as well as between two {DEM} plates,
\begin{align*}
	\mathcal{F}^{\text{PMC--DEM}}_{\text{Cas}}(L) = \mathcal{F}_{\text{Cas}}(L, \theta = 0) = -\frac{3}{8 \pi^2 L^4} \mathrm{Re} \, \mathrm{Li}_4(1) = -\frac{\pi^2}{240 L^4} = \mathcal{F}^{\text{PMC--PMC}}_{\text{Cas}}(L) = \mathcal{F}^{\text{DEM--DEM}}_{\text{Cas}}(L).
\end{align*}
On the other hand, for $\theta = \frac{\pi}{2}$, we are considering the {PEC}--{DEM} setup, and we find the same repulsive force as for the {PEC}--{PMC} setup,
\begin{align*}
	\mathcal{F}^{\text{PEC--DEM}}_{\text{Cas}}(L) = \mathcal{F}_{\text{Cas}}\left(L, \theta = \frac{\pi}{2}\right) = -\frac{7}{8} \mathcal{F}_{\text{Cas}}(L, \theta = 0) = \frac{7 \pi^2}{1920 L^4} = \mathcal{F}^{\text{PEC--PMC}}_{\text{Cas}}(L).
\end{align*}
Defining the relative Casimir force with respect to the standard Casimir force between two perfect conductors as
\begin{align}\label{eq:F-tilde}
	\tilde{\mathcal{F}}_{\text{Cas}}(\theta) \coloneqq \frac{\mathcal{F}_{\text{Cas}}(L, \theta)}{\mathcal{F}_{\text{Cas}}(L, 0)} = \frac{90}{\pi^4} \, \mathrm{Re} \, \mathrm{Li}_4(e^{2i\theta}) = \frac{90}{\pi^4} \, \sum_{n=1}^{\infty} \frac{\cos(2n\theta)}{n^4} = 1 - \frac{30}{\pi^2} \, \theta^2 + \frac{60}{\pi^3} \, |\theta|^3 - \frac{30}{\pi^2} \, \theta^4,
\end{align}
shows the dependence on the duality angle $\theta$ of the {PEMC} plate, see Fig~\ref{fig: Relative Casimir PEMC-DEM}. From this it can be seen that the {PEC}--{DEM} setup ($\theta = \frac{\pi}{2}$) is the configuration with maximal repulsion, while the {PMC}--{DEM} configuration ($\theta = 0$) is characterized by maximal attraction.

Now we can compare the Casimir force \eqref{eq: Casimir force PEMC-DEM} for the PEMC-DEM case with the force between two {PEMC} plates in Eq.~\eqref{eq: Casimir force PEMC PEMC}.
One immediately sees that the {PEMC}--{DEM} Casimir force with duality angle \(\theta\) is the same as the {PEMC}--{PEMC} Casimir force, provided that we identify $\theta$ with the difference in duality angles of the two PEMC plates: $\theta \equiv \theta^+ - \theta^-$.

Remembering how {DEM} boundary conditions were introduced, see Eq.~\eqref{eq: DEM conditions}, we see that these closely resemble {PMC} boundary conditions: they only differ in their time component. Moreover, the Casimir force in the {DEM}--{DEM} configuration, Eq.~\eqref{eq: Casimir force DEM DEM}, is given by the same expression as the Casimir force between two {PMC} plates, Eq.~\eqref{eq:PEC-PEC}. All of this suggests that, with respect to the Casimir force, a plate satisfying {DEM} conditions effectively behaves as if it were a PMC plate. Indeed, setting one of the duality angles zero and denoting the other by $\theta$ in the {PEMC}--{PEMC} setup yields exactly the same Casimir force as the {PEMC}--{DEM} configuration.

\section{Conclusion and Outlook}\label{sec:conclusion}
Following earlier work on the Casimir effect between two parallel plates satisfying {DEM} boundary conditions \cite{canforaDynamicalEdgeModes2025}, we focused here on deriving the Casimir force between a {PEMC} and {DEM} plate.  Setting $\theta = 0$ for the duality angle of the {PEMC} plate, the boundary condition reduces to a {PMC} one and gives us the Casimir force
\begin{align*}
	\mathcal{F}^{\text{PMC--DEM}}_\text{Cas}(L) = -\frac{\pi^2}{240 L^4},
\end{align*}
which is equal to the standard Casimir force between two {PEC} or two {PMC} plates, and corresponds to the case of maximal attraction. On the other hand, maximal repulsion is found for $\theta = \frac{\pi}{2}$: this is the {PEC}--{DEM} case with Casimir force ${\mathcal{F}^{\text{PEC--DEM}}_{Cas}(L) = -\frac{7}{8}\mathcal{F}^{\text{PMC--DEM}}_{Cas}(L)}$. In general, we found that the Casimir force in the {PEMC}--{DEM} configuration is identical to the Casimir force in a {PEMC}--{PMC} setup. Consistent with the {DEM}--{DEM} case, this leads us to conclude that the {DEM} plate effectively behaves as a {PMC} plate under the Casimir effect.

In order to ensure a proper BRST symmetric quantization of the {PEMC}--{DEM} action, we have introduced the field $\gamma(\bvec{x})$ living on the {DEM} plate. From the {DEM}--{DEM} case \cite{canforaDynamicalEdgeModes2025}, we know that this field is needed to arrive at a gauge-independent Casimir energy. In the {PEMC}--{DEM} case, if we had not introduced $\gamma(\bvec{x})$, the Casimir energy would be given in terms of $|\mathbb{M}| = |\mathbf{A}| \cdot |\mathbb{M}/\mathbf{A}|$ instead of $|\mathbf{A}|$, and the factor $|\mathbb{M}/\mathbf{A}|$ would have introduced a $\xi$-dependence in the Casimir energy and force.  We verified explicitly that the Casimir energy and force found in Landau and Coulomb gauge are identical.

Consequently, the Casimir energy only depends on the $5\times5$ submatrix $\mathbf{A}$ of $\mathbb{M}$. In other words, the dynamical edge modes, $b_t^+$, do not contribute to the Casimir energy and force. Comparing the calculations in the {PEMC}--{PMC} case with the calculation in the {PEMC}-{DEM} setups, the only difference lies in the role of the Lagrange multiplier field $b_t^+$, which imposes different conditions on the {PMC}, respectively {DEM}, plate,
\begin{align*}
	(\text{{PMC}}) \quad & F_{zt} = 0, & (\text{{DEM}}) \quad & A_t = 0.
\end{align*}
In the {PEMC}--{DEM} calculation, the $\gamma(\bvec{x})$ field cancels the contribution to the Casimir energy of the $b_t^+$-dependent terms in the action. On the other hand, in the {PEMC}--{PMC} case, we can make use of the residual gauge symmetry $b_i^-(\cev{k})\rightarrow b_i^-(\cev{k}) + k_i \chi(\cev{k})$ to set $b_t^+ = 0$. As a consequence, in the Casimir energy calculation, the determinant of the quadratic operator reduces to the $5\times5$ submatrix $\mathbf{A}$,
\begin{align*}
	|\mathbb{M}| \xrightarrow{\text{Choose: } b_t^+ = 0} |\mathbf{A}|.
\end{align*}
This explains why the {DEM} plate behaves as if it were a {PMC} plate under the Casimir effect.

An interesting way to extend this work would be to add a chiral medium in between the {PEMC} and {DEM} plates (or between two {DEM} plates) \cite{canforaCasimirEffectChiral2022,oosthuyseInterplayChiralMedia2023}. Since we then lose the residual gauge symmetry in the $b$-fields, it is possible for the {DEM} plate to behave in a non-trivial way, yielding a different Casimir force.

In addition, one could consider these {PEMC} and {DEM} boundary conditions in other configurations than the parallel plate setup or even work in different, more general, spacetimes than the flat Euclidean/Minkowski spacetime manifold we used here. Such curved manifolds might be of interest to the gravitational community.

Finally, connecting more directly with possible real-world applications, it would be interesting to see if and how these {DEM} boundary conditions could be realized in real materials.

\section*{Acknowledgments}
The work of D.D.~was supported by KU Leuven IF project C14/21/087. The work of S.S.~was funded by FWO PhD-fellowship fundamental research (file number: 1132823N).

\appendix

\section{Useful integrals}\label{sec: useful fourier integrals}
In the calculation of the Casimir energy we often encounter Fourier integrals over the $z$-component of the wavevector, $k_z$, which is the direction orthogonal to the parallel plates. In this appendix we present a list of integrals of interest.

All of the integrals we need can be obtained from the integral,
\begin{align}
	I(a,y,s) & = \int_{-\infty}^{\infty} \frac{\mathrm{d}x}{2\pi} \frac{e^{\pm i x a}}{(x^2+y^2)^s} \text{ with } y>0, \, a>0 \nonumber \\
	& = 2\int_{0}^{\infty} \frac{\mathrm{d}x}{2\pi} \cos(a x)(x^2+y^2)^{-s} \nonumber \\
	& = 2^{\frac{1}{2}-s} \left(\frac{a}{y}\right)^{s-\frac{1}{2}}\frac{K_{s-\frac{1}{2}}(ay)}{\sqrt{\pi} \Gamma(s)} \label{eq: fund integral}
\end{align}
which is convergent for $\mathrm{Re}(s)>0$. In going to the second line we dropped the term $\pm i \sin(ax)(x^2+y^2)^{-s}$ since it is an odd function of $x$ and integrates to zero. The last line can be verified using expressions for Fourier cosine transforms in integration tables such as \cite{gradshteinTableIntegralsSeries2015}. Here, $K_{\alpha}(x)$ is the modified Bessel function of the second kind and $\Gamma(s)$ is the Gamma function. Using that $K_{1/2}(x) = \sqrt{\frac{\pi}{2x}}e^{-x}, K_{3/2}(x) = \sqrt{\frac{\pi}{2x}}e^{-x}(1+\frac{1}{x})$ and $k^2 = |\bvec{k}|^2 + k_z^2$ we find the following (regularized) integrals
\begin{align}
	&\int \frac{\mathrm{d}k_z}{2\pi} \frac{e^{\pm i k_z a}}{k^2} = I(a,|\bvec{k}|,1) = \frac{e^{-|\bvec{k}| a}}{2|\bvec{k}|} & \xrightarrow{\ a \, \rightarrow \, 0 \ } && \int \frac{\mathrm{d}k_z}{2\pi} \frac{1}{k^2} & = \frac{1}{2|\bvec{k}|} \label{eq: kz0 k2} \\
	&\int \frac{\mathrm{d}k_z}{2\pi} \frac{k_z^2 e^{\pm i k_z a}}{k^2} = -\frac{d^2 I(a,|\bvec{k}|,1) }{da^2} = -\frac{|\bvec{k}|}{2}e^{-|\bvec{k}| a} & \xrightarrow{\ a \, \rightarrow \, 0 \ } && \int \frac{\mathrm{d}k_z}{2\pi} \frac{k_z^2}{k^2} & = -\frac{|\bvec{k}|}{2} \label{eq: kz2 k2}\\
	&\int \frac{\mathrm{d}k_z}{2\pi} \frac{e^{\pm i k_z a}}{k^4}  = I(a,|\bvec{k}|,2) = \frac{1+|\bvec{k}| a}{4|\bvec{k}|^3}e^{-|\bvec{k}| a} & \xrightarrow{\ a \, \rightarrow \, 0 \ } && \int \frac{\mathrm{d}k_z}{2\pi} \frac{1}{k^4} & = \frac{1}{4|\bvec{k}|^3} \label{eq: kz0 k4} \\
	&\int \frac{\mathrm{d}k_z}{2\pi} \frac{k_z^2 e^{\pm i k_z a}}{k^4} = -\frac{d^2 I(a,|\bvec{k}|,2) }{da^2} = \frac{1-|\bvec{k}| a}{4|\bvec{k}|}e^{-|\bvec{k}| a} & \xrightarrow{\ a \, \rightarrow \, 0 \ } && \int \frac{\mathrm{d}k_z}{2\pi} \frac{k_z^2}{k^4} & = \frac{1}{4|\bvec{k}|}\label{eq: kz2 k4}\\
	&\int \frac{\mathrm{d}k_z}{2\pi} \frac{k_z e^{\pm i k_z a}}{k^2} = \mp i \frac{d I(a,|\bvec{k}|,1) }{da} = \pm \frac{i}{2}e^{-|\bvec{k}| a} & \xrightarrow{\ a \, \rightarrow \, 0 \ } && \int \frac{\mathrm{d}k_z}{2\pi} \frac{k_z}{k^2} & = 0. \label{eq: kz1 k2}
\end{align}

\section{Block matrices and the Schur complement}\label{ch: Schur}
When computing the Casimir energy between a {PEMC} and a {DEM} plate, the calculation involves the submatrix $\left(\mathbb{M}(\bvec{k})^{-1}\right)_{\mathbf{tt}}^{++}$ of the full matrix inverse $\mathbb{M}(\bvec{k})^{-1}$, see equation \eqref{eq: part function gamma Landau}. A useful mathematical tool in this context is the Schur complement, which can help us with the calculation of determinants and matrix inverses of block matrices.

To introduce the Schur complement we start from a general $n \times n$ matrix $M$ that can be written as a $2\times2$ block matrix
\begin{align*}
	M = \left(
	\begin{array}{cc}
		A_{p \times p} & B_{p \times q}\\
		C_{q \times p} & D_{q \times q}
	\end{array}
	\right)
\end{align*}
where $A$ and $D$ are square $p \times p$ and $q \times q$ matrices respectively with $n = p + q$. When the square matrix $A$ is invertible, one can write the matrix $M$ in a lower-diagonal-upper (LDU) decomposition\footnote{Here we understand diagonal in this context as being a block diagonal matrix.}
\begin{align*}
	M = \left(
	\begin{array}{cc}
		\mathbbm{1}_p & 0\\
		C A^{-1} & \mathbbm{1}_q
	\end{array}
	\right)
	\cdot
	\left(
	\begin{array}{cc}
		A & 0\\
		0 & M/A
	\end{array}
	\right)
	\cdot
	\left(
	\begin{array}{cc}
		\mathbbm{1}_p & A^{-1}B\\
		0 & \mathbbm{1}_q
	\end{array}
	\right),
\end{align*}
with the Schur complement $M/A$, defined by
\begin{align*}
	M/A \coloneqq D-C A^{-1} B.
\end{align*}

From the {LDU} decomposition for the block matrix $M$ above, if the Schur complement $M/A$ is invertible, we can immediately find the inverse of $M$ as
\begin{align}
	M^{-1} & =
	\left(
	\begin{array}{cc}
		\mathbbm{1}_p & A^{-1}B\\
		0 & \mathbbm{1}_q
	\end{array}
	\right)^{-1}
	\cdot
	\left(
	\begin{array}{cc}
		A & 0\\
		0 & M/A
	\end{array}
	\right)^{-1}
	\cdot
	\left(
	\begin{array}{cc}
		\mathbbm{1}_p & 0\\
		C A^{-1} & \mathbbm{1}_q
	\end{array}
	\right)^{-1} \nonumber \\
	& =
	\left(
	\begin{array}{cc}
		\mathbbm{1}_p & -A^{-1}B\\
		0 & \mathbbm{1}_q
	\end{array}
	\right)
	\cdot
	\left(
	\begin{array}{cc}
		A^{-1} & 0\\
		0 & (M/A)^{-1}
	\end{array}
	\right)
	\cdot
	\left(
	\begin{array}{cc}
		\mathbbm{1}_p & 0\\
		-C A^{-1} & \mathbbm{1}_q
	\end{array}
	\right)  \nonumber \\
	& = \left(
	\begin{array}{cc}
		A^{-1} + A^{-1}B(M/A)^{-1}CA^{-1} & -A^{-1}B(M/A)^-1\\
		-(M/A)^{-1}CA^{-1} & (M/A)^{-1}
	\end{array}
	\right). \label{eq: Schur inverse}
\end{align}
As a consequence, when only a submatrix of $M^{-1}$ is required in a calculation, it is not needed to compute the full matrix inverse of $M$. Instead one can check that both $A$ and its Schur complement $M/A$ are invertible, and then with the above expression determine the desired submatrices of the inverse.

Another useful result from the above {LDU} decomposition, using the Schur complement $M/A$, is the determinant of $M$. Since both the lower and upper triangular matrix have entries one along its diagonal, their determinants are simply one. Hence the determinant of $M$ is given by the determinant of the block diagonal matrix
\begin{align}
	|M| =
	\left|
	\begin{array}{cc}
		A & 0\\
		0 & M/A
	\end{array}\right|
	= |A| \cdot |M/A|, \label{eq: Schur determinant}
\end{align}
where $|\dots|$ denotes the matrix determinant.

\bibliographystyle{apsrev4-2}
\bibliography{bib}

\end{document}